\documentclass[fleqn,usenatbib]{mnras}

\usepackage{mathptmx}

\usepackage[T1]{fontenc}

\DeclareRobustCommand{\VAN}[3]{#2}
\let\VANthebibliography\thebibliography
\def\thebibliography{\DeclareRobustCommand{\VAN}[3]{##3}\VANthebibliography}

\usepackage{graphicx}	
\usepackage{amsmath}	
\usepackage{amssymb}	
\usepackage{xspace}
\usepackage{longtable}
\usepackage{subfig}
\usepackage{booktabs}
\usepackage{lscape}
\usepackage{siunitx}
\usepackage[symbol]{footmisc}
\usepackage[dvipsnames]{xcolor}

\newcommand{\Me}{\mathrm{M}_\oplus}

\newcommand{\tess}{\ensuremath{\emph{TESS}}\xspace}

\newcommand{\gaia}{\textit{Gaia}\xspace}
\newcommand{\ariel}{\ensuremath{\emph{Ariel}}\xspace}
\newcommand{\paws}{\texttt{PAWS}\xspace}

\newcommand{\ispec}{\texttt{iSpec}\xspace}

\newcommand{\teff}{\ensuremath{T_{\textup{eff}}}\xspace}
\newcommand{\feh}{\ensuremath{[\text{Fe/H}]}\xspace}

\newcommand{\review}[1]{{#1}}

\title[Two very low-density planets around TOI-791]{ASTEP confirmation of \review{a pair of long-period Jupiter-sized} planets \review{with extremely low densities} transiting TOI-791}

\author[Georgina Dransfield et.al.]{
Georgina Dransfield$^{1,2,3}$,
Antoine C. Petit$^{4}$,
Amaury H.M.J. Triaud$^{3}$,
Tristan Guillot$^{4}$,
\newauthor
Fran{\c c}ois-Xavier Schmider$^{4}$,
Lyu Abe$^{4}$,
Abdelkrim Agabi$^{4}$,
Khalid Barkaoui$^{5,6,7}$,
Thomas A. Baycroft$^{3}$,
\newauthor
Philippe Bendjoya$^{4}$,
Rafael Brahm$^{8,9,10}$,
Karen A. Collins$^{11}$,
Billy Edwards$^{12}$,
Phil Evans$^{13}$,
\newauthor
Alix V. Freckelton$^{3}$,
Nolan Grieves$^{14}$,
Steve B. Howell$^{15}$,
Franco Mallia$^{16}$,
Djamel Mekarnia$^{4}$,
\newauthor
Angelica Psaridi$^{14,17,18}$,
Daniel Sebastian$^{19,3}$,
Keivan G. Stassun$^{20}$,
Chris Stockdale$^{21}$,
Amalie Stokholm$^{3}$,
\newauthor
Olga Suarez$^{4}$,
Thiam-Guan Tan$^{22}$,
Mathilde Timmermans$^{3,6}$,
Cristilyn N. Watkins$^{11}$,
Carl Ziegler$^{23}$,
\newauthor
Abderahmane Soubkiou$^{6}$,
Fran{\c c}ois Bouchy$^{14}$,
Marion Cointepas$^{14,24}$,
Vincent Deloupy$^{4}$,
\newauthor
Maximilian N. G{\"u}nther$^{25}$,
Micha{\"e}l Gillon$^{6}$,
Giovanni Isopi$^{26,27,28,29}$,
Emmanuel Jehin$^{30}$,
Jon M. Jenkins$^{15}$,
\newauthor
Andr\'es Jord\'an$^{8,9,10}$,
Martin B. Nielsen$^{3}$,
Sara Seager$^{7,31,32}$,
Avi Shporer$^{7}$,
Julia V. Seidel$^{4,14,33}$,
\newauthor
Michal Steiner$^{14}$,
Trifon Trifonov$^{34,35}$,
Joseph D. Twicken$^{15,36}$,
Joshua N. Winn$^{37}$,
Aldo Zapparata$^{16}$
\newauthor
\\
The list of affiliations is given at the end of the paper
}
\date{Accepted 2026 May 06. Received 2026 April 23; in original form 2024 December 12}

\pubyear{2025}

\begin{document}
\label{firstpage}
\pagerange{\pageref{firstpage}--\pageref{lastpage}}
\maketitle

\begin{abstract}
Gas giant planets with periods $20~<~P~<~300~\rm days$ orbiting Sun-like stars are a relatively uncommon outcome of planetary formation, and key questions about the nature and formation of this sub-population remain unanswered. Theoretical models for the location of their formation (in- or ex-situ) and for their subsequent migration predict different outcomes in terms of planet masses and eccentricities, indicating that observations have a key role to play in disentangling their histories. In this work we present the discovery and confirmation of a pair of long-period Jupiter-sized planets transiting an F7 star: TOI-791~b is a $0.993\pm0.033\rm~R_{Jup}$ planet on a $139.29931_{-0.00012}^{+0.00011}~\rm day$ orbit, and TOI-791~c, a $1.155\pm0.040\rm ~R_{Jup}$ planet on a $232.01570_{-0.00071}^{+0.00067}~\rm day$ orbit. The two planets are within 0.07\% of a second-order 5:3 period commensurability leading to transit timing variations (TTVs) of up to \review{50} minutes. We confirm their planetary nature using ground-based photometry, including multiple full detections of the $>11~\rm hr$ transits of \review{both} TOI-791~b \review{and c} from Antarctica with ASTEP, making these the longest-duration transits ever observed in their entirety from the ground. \review{Our detailed analysis of the TTV signal allows us to measure dynamical masses for both planets, which yield densities of $\rho_{\rm b}=0.038\pm0.008 \rm ~g~cm^{-3}$ and $\rho_{\rm c}=0.047\pm0.006 \rm ~g~cm^{-3}$, indicating that TOI-791~b and c are two of the lowest density giant planets ever detected. {While these measurements are robust, further follow-up is needed to fully characterise the TTV signal and the architecture of the system}.}


\end{abstract}




\section{Introduction}


A transiting warm Jupiter is a rare beast. According to the NASA Exoplanet Archive \citep{Akeson2013}, $\sim30\%$ of the known planet population (\review{1960 out of 6291} planets) are giant planets with radii $\geq 0.7~\rm R_{Jup}$; of these \review{261} have orbital periods in the range $20~<~P~<~300~\rm days$. However, fewer than half of them are known to transit, meaning that \review{$\sim 1.5\%$} of planets discovered to date are transiting warm Jupiters\footnote{Figures correct as of \review{August 2025}.}. 

Giant planet formation remains shrouded in mystery, despite a vast catalogue of investigations into the hot Jupiter population spanning two and a half decades. Hot Jupiters are by far the easiest kind of planets to study, producing large signals over short timescales in photometry, radial velocity, and transmission spectroscopy. There is consensus that hot Jupiters most likely form by core accretion \citep{Pollack1996,Chabrier2014}, as gravitational instability has not proven feasible in numerical simulations \citep{Stamatellos2008,Dawson2018}. Core accretion could take place `in situ' \citep[e.g.][]{Batygin2016} or further out and be followed by migration \citep[e.g.][]{Ida+Lin2004, Alibert2005}, although the former is thought to be less likely due to the large amount of solids needed in the disc \citep{Rafikov2006,Piso2015}. The mechanism by which these nascent planets migrate is also contested, the key options being disc migration \citep{Lin1996,Baruteau2014} or high eccentricity tidal migration \citep{Rasio1996}. 

Occurrence rate studies indicate that warm Jupiters are less abundant than their cold ($P>300~\rm days$) counterparts \citep{Santerne2016, Winn2018, Wittenmyer2020, Su2024}, and it has been suggested that some members of this intermediate population could be hot Jupiter progenitors caught in the act of inward migration \citep[e.g.][]{Dawson2015,Dong2021}. If this is the case, they are the bridge between the well-characterised hot Jupiters and the elusive largely non-transiting cold Jupiters, making them a keystone population to probe giant planet formation as a whole.

Confronting formation theory with observation requires the detection of \textit{transiting} \review{giant planets} specifically, as precise mass and radius constraints are needed for bulk density calculations and interior modeling \citep{Guillot1999,Thorngren2019, Guillot+2023PP7}. Additionally, a transiting configuration is necessary for other key investigations such as transmission spectroscopy and obliquity measurements via the Rossiter-McLaughlin effect, both of which can give hints about a planet's origin story \citep{Madhusudhan2014, Triaud2018}. It is timely then that \textit{TESS} \citep{tess} has been providing new warm Jupiter discoveries to the community since its launch in 2018. 

While \textit{TESS}'s 27-day sectors are not necessarily conducive to a bumper crop of long-period planetary candidates, areas of overlap (in particular the continuous viewing zones) are. The challenge then lies in transforming candidates into fully-fledged planets, as long periods often also imply long transit durations, both of which present observational challenges for  most ground-based telescopes. In cases like these, one southern telescope excels: ASTEP \citep[the Antarctic Search for Transiting ExoPlanets;][]{astep400,Guillot2015,astep+} is a 40~cm optical telescope located on the Antarctic plateau, where it enjoys nearly uninterrupted observing during the Austral winter. This means that ASTEP is uniquely able to observe transits that are both long-duration and infrequent \citep{ASTEP}, leading to the discovery of numerous long-period transiting systems \citep[e.g.][]{TOI282,trifonov2023,Hobson2023}. 

Almost $40\%$ of known warm Jupiters, transiting or otherwise, are accompanied by at least one other known planet \citep[e.g.][]{Huang2016}, but even more of a unicorn than a transiting warm Jupiter is a system containing multiple transiting warm Jupiters. In fact, only 8 systems are known to contain more than one transiting giant planet, and in every case the giants are dynamically interacting and present transit timing variations (TTVs){; they are: Kepler-9~b~\&~c, \cite{kepler9}; Kepler-30~c~\&~d, \cite{kepler30}; Kepler-51~c~\&~d, \cite{kepler51}; Kepler-108~b~\&~c and Kepler-117~b~\&~c, \cite{kepler108}; Kepler-1513~b~\&~c, \cite{kepler1513}; Kepler-90~g~\&~h, \cite{koi351}; and TOI-2525~b~\&~c, \cite{trifonov2023}}. \review{Of the planets here listed, Kepler-51~c~\&~d are unique: their large radii and orbital periods place them in the warm Jupiter regime, yet their densities are among the lowest ever measured, meaning they are also classified as `super-puffs'. }{As is common for multi-transiting super-puff systems, the masses of the Kepler-51 planets were measured using TTVs.}

In this context, we expand this \review{very} short list by reporting on the discovery of TOI-791~b and c, a pair of long-period, dynamically interacting, transiting \review{Jupiter-sized planets with extremely low densities} orbiting an F7 dwarf.

Our paper begins with in-depth characterisation of the host star in Section \ref{sec:star}, followed by the identification of planetary candidates in Section \ref{sec:sherlock}. Section \ref{sec:followup} presents our ground-based follow-up campaign, and modeling of all the available data is presented in Section \ref{sec:analysis}. \review{In Section \ref{sec:mass_limits} we present measurements of the planets' masses.} We contextualise the system in Section \ref{sec:discussion}, and finally conclude in Section \ref{sec:conc}.

\section{Stellar Characterisation}
\label{sec:star}

TOI-791 is an F7-type star in \textit{TESS}'s southern Continuous Viewing Zone (CVZ). Its right ascension is 07:53:44.34 hms and its declination is -69:42:06.25 dms.

As all our planetary information will be derived using the host star's parameters, we begin in the following sections by characterising TOI-791. All photometry and stellar parameters adopted for this work can be found in Table \ref{tab:starpar}.

\begin{table}
\centering
\caption{Stellar parameters adopted for this work.}
\begin{tabular}{@{}lp{25mm}p{30mm}@{}}
\toprule
{\bf Designations} & \multicolumn{2}{p{65mm}}{TOI-791, TIC 306472057, 2MASS J07534434-6942063, APASS 33130674, Gaia DR2 5270467966514084224, TYC 9184-01675-1, UCAC4 102-020935, WISE J075344.34-694206.2} \\ \midrule
{\bf Parameter} & {\bf Value}              & {\bf Source} \\ \midrule
T mag           & 10.753$\pm$0.006        & \cite{TICv8} \\
B mag           & 11.368$\pm$0.127      & \cite{ucac4} \\
V mag           & 11.09$\pm$0.01        & \cite{ucac4} \\
G mag           & 11.0873$\pm$0.0005      & \cite{gaiaDR3cat} \\
J mag           & 10.269$\pm$0.023          & \cite{2masscat} \\
H mag           & 10.061$\pm$0.022          & \cite{2masscat} \\
K mag           & 9.992$\pm$0.023         & \cite{2masscat} \\
W1 mag           & 9.939$\pm$0.023          & \cite{wisecat} \\
W2 mag           & 9.95$\pm$0.02          & \cite{wisecat} \\
W3 mag           & 9.878$\pm$0.028            & \cite{wisecat} \\
W4 mag           & 9.269$\pm$0.262       & \cite{wisecat} \\
Distance         & 341.221$\pm$3.074\,pc       & \cite{BJdist} \\
$\alpha$           & 07:53:44.34      & \cite{gaiaDR3cat} \\
$\delta$           & -69:42:06.25      & \cite{gaiaDR3cat} \\
$\mu_{\alpha}$           & $\rm -5.139\,mas\,yr^{-1}$      & \cite{gaiaDR3cat} \\
$\mu_{\delta}$           & $\rm 8.285\,mas\,yr^{-1}$      & \cite{gaiaDR3cat} \\
SpT             & F7                  & This work \\
$R_{\star}$     & $1.474\pm 0.047~\rm R_{\odot}$ & This work                  \\
$M_{\star}$     & $1.28\pm 0.08~\rm M_{\odot}$   & This work                  \\
${\rm T_{eff}}$ & 6294$\pm$128 K             & This work                  \\
$\log g_\star$           & 3.71$\pm$0.36           & This work                  \\
$\rm [Fe/H]$        & -0.14$\pm$0.14 dex              & This work (spectroscopy)   \\ 
$v\sin i$     &     44.58$\pm$2.28 $\rm km~s^{-1}$    & This work    \\
$\rm RV$    &     6.238$\pm$0.949 $\rm km~s^{-1}$  &   \cite{gaiaDR3cat} \\
         
\bottomrule
\end{tabular}
\label{tab:starpar}
\end{table}

\subsection{High-resolution spectroscopy}
\label{sec:recspec}

We collected high resolution spectra with both CORALIE and FEROS to characterise TOI-791 and check for the presence of companions. We describe these observations in the following two sections.

\subsubsection{CORALIE}


TOI-791 was observed with the CORALIE spectrograph on the Swiss 1.2-m telescope at La Silla, Chile \citep{Queloz01} between November 2019 and February 2022. A total of 15 spectra were obtained. The data were reduced using an adapted version of the HARPS standard data reduction pipeline. For each epoch, we derived the radial velocity (RV) of the star using the cross-correlation function (CCF) technique \citep{pepe2002} using the A0 mask \review{as it gave the best match}. We also computed CCFs with other masks to check for the presence a secondary component in the spectra. The observations showed RV variations at the level of $380~\rm m~s^{-1}$ RMS, we thus ruled out double-lined binaries, as well as companions with a mass greater than $13~{\rm M_{jup}}$ at $3-\sigma$. The RV measurements can be found in Table \ref{tab:coralie}

Spectra were then shifted into the lab frame using a CCF to identify the necessary radial velocity correction. All available spectra were coadded prior to stellar parameter extraction to achieve a SNR of 262. Stellar parameters were determined using the \paws pipeline, as described by \citet{freckelton2024}. The curve-of-growth equivalent widths method was skipped in this instance due to the projected rotational velocity $v\sin i > 5 ~\rm km~s^{-1}$, as measured from the CCF. \citet{tsantaki2014} describe how the performance of the equivalent widths method is reduced for fast rotating stars due to the blending of spectral lines. Instead of initial estimates being from the equivalent widths step, as is the usual process in the \paws pipeline, the standard initial estimates for an F type star were used : ${\rm T_{eff}}$ = 6750 K, $\log g_\star$ = 4.2 dex, $\rm [Fe/H]$ = 0.00 dex.  These initial estimates were used to start the spectral synthesis, implemented via the \ispec package  \citep{BlancoCuaresma2014, BlancoCuaresma2019} using the \texttt{SPECTRUM} \citep{gray1994} line list and the \texttt{ATLAS} \citep{kurucz2005} set of model atmospheres to produce the final stellar atmospheric parameters displayed in Table \ref{tab:starpar}.

\subsubsection{FEROS}
 
We obtained three spectra of TOI-791 with the FEROS instrument \citep{feros} installed at the MPG 2.2~m telescope at the ESO La Silla Observatory in Chile. FEROS is a fibre-fed high resolution ($R=48,000$) \'echelle spectrograph with a wide wavelength coverage. FEROS contains a second fibre that can be used simultaneously during the science exposure to trace instrumental velocity variations illuminating it with a ThAr lamp. The FEROS observations were executed between June of 2019 and March of 2020 in the context of the Warm gIaNts with tEss (WINE) collaboration \citep{wine0,wine1,wine2,wine3,wine4,wine5,wine6,trifonov2023,wine8}. We adopted exposure times between 600~s and 900~s that generated spectra with signal-to-noise ratios per resolution element between 40 and 80. FEROS data were processed from raw images using the automated pipeline built with the \textsc{ceres} \citep{ceres} package to obtain the final spectra and additional outputs, including precision radial velocities, bisector span measurements and a rough estimation of the stellar atmospheric parameters. We measured a  projected rotation velocity for TOI-791 of $v\sin i = 43 \pm 3 \rm km~s^{-1}$. Such a relatively high level of Doppler broadening led to large uncertainties ($200 - 300\ \rm m~s^{-1}$) in the radial velocity measurements. No significant radial velocity variations were identified from the three FEROS radial velocity measurements.

\subsection{Spectral Energy Distribution}
\label{sec:sed}

\begin{figure}
    \centering
    \includegraphics[width=0.7\linewidth,trim=70 80 80 100,clip,angle=90]{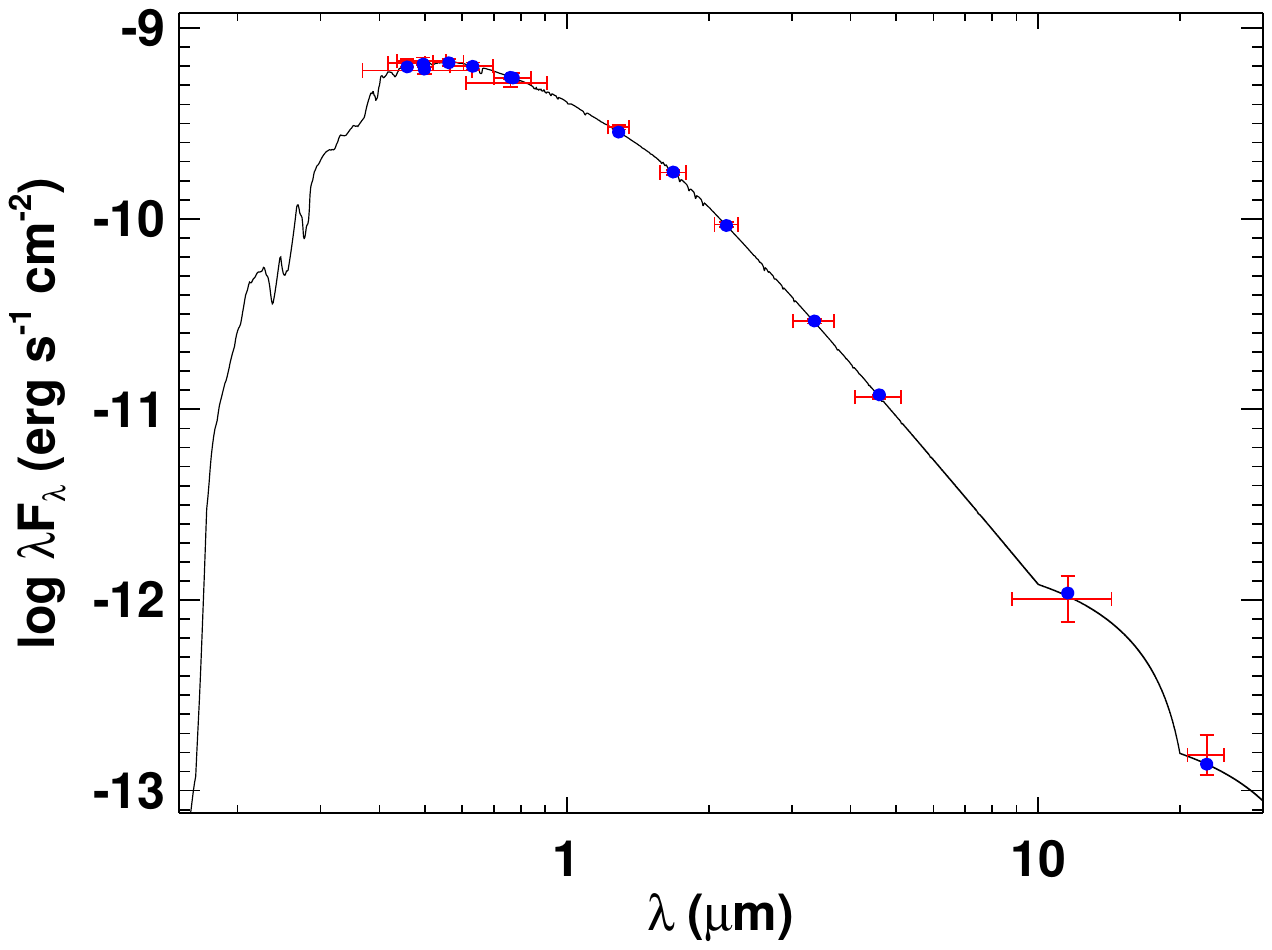}
\caption{Spectral energy distribution of TOI-791. Red symbols represent the observed photometric measurements, where the horizontal bars represent the effective width of the passband. Blue symbols are the model fluxes from the best-fit Kurucz atmosphere model (black).  \label{fig:sed}}
\end{figure}

As an independent determination of the basic stellar parameters, we performed an analysis of the broadband spectral energy distribution (SED) of the star together with the {\it Gaia\/} DR3 parallax \citep[with no systematic offset applied; see, e.g.,][]{StassunTorres:2021}, in order to determine an empirical measurement of the stellar radius, following the procedures described in \citet{Stassun:2016,Stassun:2017,Stassun:2018}. We pulled the $BVgri$ magnitudes from {\it APASS}, the $JHK_S$ magnitudes from {\it 2MASS}, the W1--W4 magnitudes from {\it WISE}, and the $G_{\rm BP} G_{\rm RP}$ magnitudes from {\it Gaia}. Together, the available photometry spans the full stellar SED over the wavelength range 0.4--20~$\mu$m (see Figure~\ref{fig:sed}).  

We performed a fit using Kurucz stellar atmosphere models, with the effective temperature ($\teff$) and metallicity (\feh) adopted from the spectroscopic analysis, the remaining free parameter being the extinction $A_V$, which we limited to the maximum line-of-sight value from the Galactic dust maps of \citet{Schlegel:1998}. The resulting fit (Fig.~\ref{fig:sed}) has a best-fit $A_V = 0.11 \pm 0.02$, with a reduced $\chi^2$ of 1.0. Integrating the (unreddened) model SED gives the bolometric flux at Earth, $F_{\rm bol} = 9.57 \pm 0.11 \times 10^{-10}$ erg~s$^{-1}$~cm$^{-2}$. Taking the $F_{\rm bol}$ and $\teff$ together with the {\it Gaia\/} parallax, gives the stellar radius, $R_\star = 1.474 \pm 0.047$~R$_\odot$. In addition, we can estimate the stellar mass from the empirical relations of \citet{Torres:2010}, giving $M_\star = 1.28 \pm 0.08$~M$_\odot$. 

Finally, the spectroscopically determined $v\sin i$ together with $R_\star$ above gives an estimate of the maximum possible stellar rotation period, $P_{\rm rot}/\sin i = 1.78 \pm 0.14$ days. \review{Our understanding of the spin evolution of these stars indicates that it should be younger than $1-2~\rm Gyr$ \citep{SadeghiArdestani+2017}.}

\subsection{Stellar signature in the TESS light-curve}
We searched for evidence of asteroseismic oscillations in the \tess 2-minute cadence light-curve (described in the next section), the analysis of which would improve our ability to characterise the star. Given the effective temperature, it was tempting to look for $\delta$ Scuti photometric variations, as the temperature range of these stars is generally considered between 6000 to 9000 K \review{\citep{2011Uytterhoeven}}. However, both the estimated mass and color-luminosity relation of the star clearly exclude this possibility. From \cite{2024ApJ...972..137G}, it can be seen that the star falls outside the instability strip determined from a sample of 1000 \tess photometric variables. Therefore, only solar-type oscillations could exist, with photometric variations certainly not larger than 10~ppm -. Such oscillations would be very difficult to detect with the precision of the \tess photometric measurements. Indeed, the standard deviation of the \tess light curve for TOI-791 is 1.5 ppt for the 2-minute cadence. If the mode lifetime were infinite, \review{w}ith 3 years of data and a time coverage of 36\% as we have with TESS, it would be possible to detect modes with an amplitude of 20 ppm with a 3 $\sigma$ level \review{assuming red noise in the \tess data is not dominating in the frequency range above 100 $day^{-1}$}, where solar-type oscillations are expected. The lifetime of solar-type oscillations is usually not longer than a few days; therefore, only modes with amplitude larger than 100 ppm could be detectable.

\begin{figure}
    \centering
    \includegraphics[width=9cm]{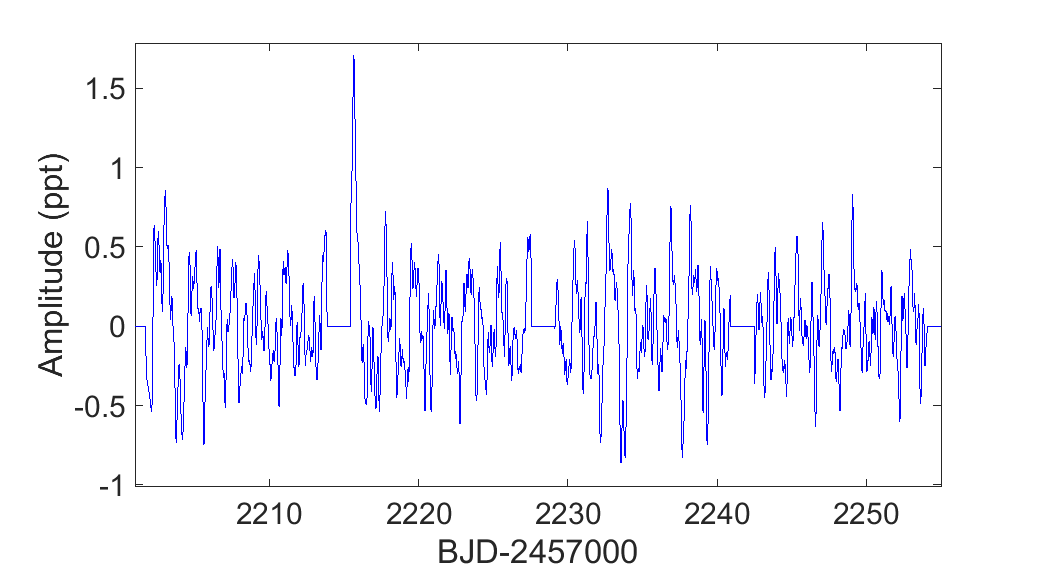}
    \caption{Detail of the TESS light-curve obtained during sectors 33 and 34. High frequency noise higher than 24 c/d (1-hour) has been filtered}
    \label{fig:sector}
    
\end{figure}

\begin{figure}
    \centering
    \includegraphics[width=9cm]{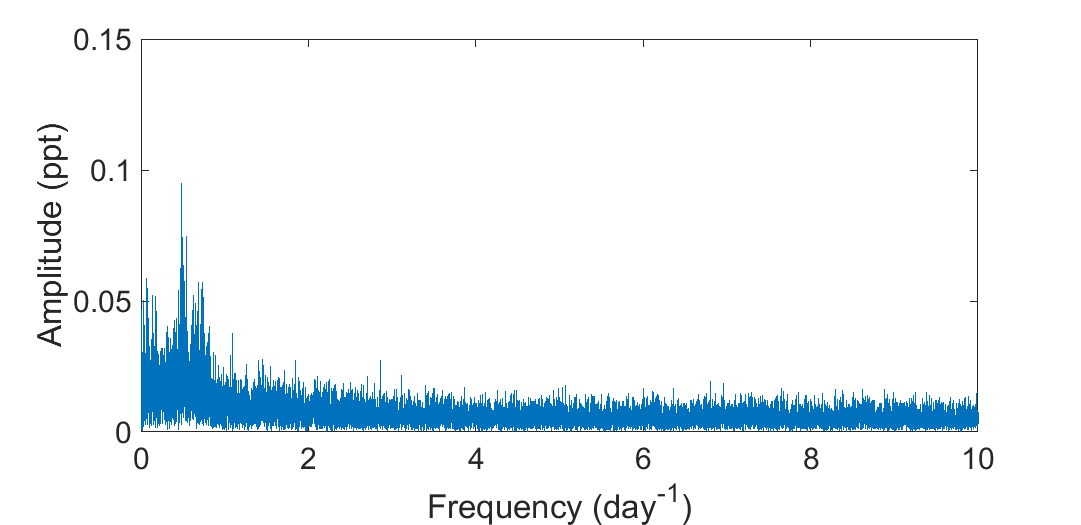}
    \caption{Density power spectrum of the three years of TESS light-curve}
    \label{fig:spectrum}
\end{figure}

Nevertheless, pseudo-periodic variations are clearly visible in the light-curve \review{as can be seen in Figure} \ref{fig:sector}, with periodicity of almost 2 days.  Two peaks are present in the power spectrum of the time series \review{as shown in Figure} \ref{fig:spectrum}, at 0.545 cycle/day and 0.72 cycle/day, with their respective harmonics at 1.09 and 1.45 cycle/day, corresponding to periods of 1.83 and 1.38 days. The same two peaks are intermittently present over the three years. The amplitude of the highest peak at 1.83 days corresponds to a variation of 100~ppm. The power spectrum shows a perfectly flat white noise above 3 cycle/day (35~$\mu$Hz). These variations could be interpreted as the signature of the stellar rotation, with spots present at the surface of the star. A 2-day rotation period is compatible with the estimated stellar diameter and the sky-projected rotational velocity of  $44~\rm km~s^{-1}$. A large number of active stars exhibit not only one, but 2 periods corresponding to the rotation, as seen with {\it Kepler} \citep{2013Reinhold}, which is interpreted as the effect of differential rotation. This is probably also the case for TOI-791. 



\section{Identification of planetary candidates}
\label{sec:sherlock}

\begin{figure*}
    \centering
    \includegraphics[width=0.85\textwidth]{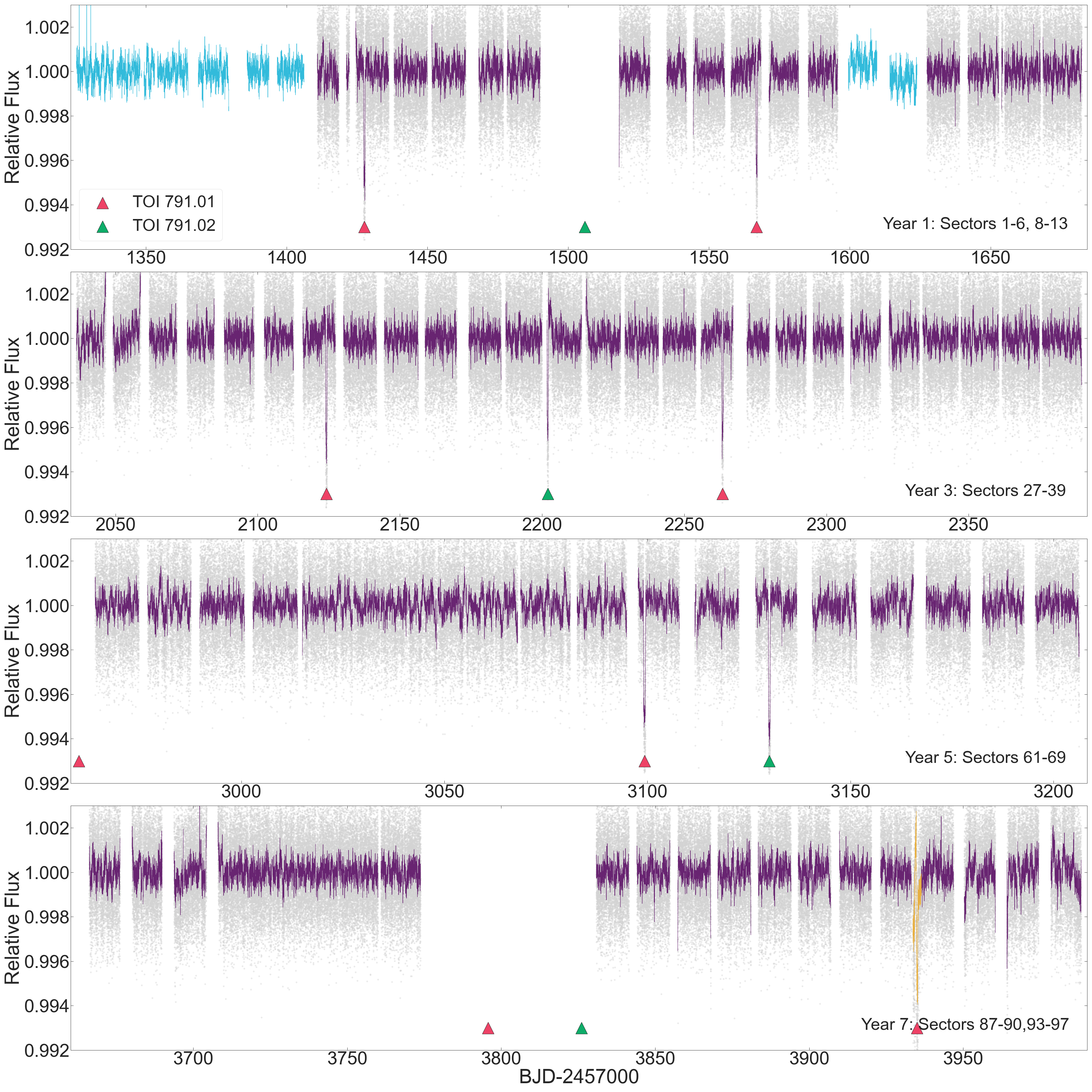}
    \caption{Short and long cadence \tess photometry for TOI-791 from 2018 to 2023; where 2~min photometry was not available, we show the 30~min photometry for that sector instead. 30~min photometry produced by the {\sc TESS-SPOC} pipeline \citep{Caldwell2020} is shown in blue in the upper panel. The PDCSAP flux as output by the {\sc SPOC} pipeline is shown in grey, with the 30~min binned points shown in purple. {In the bottom panel, a small potion of the binned lightcurve is shown in yellow to indicate that this is SAP (Simple Aperture Photometry) flux as this section of the lightcurve was not available as PDCSAP due to large systematics.} Pink and green arrows indicate the timings of transits of TOI-791.01 and TOI-791.02 respectively.}
    \label{fig:tess}
\end{figure*}

\begin{figure*}
    \centering
    \includegraphics[width=\textwidth]{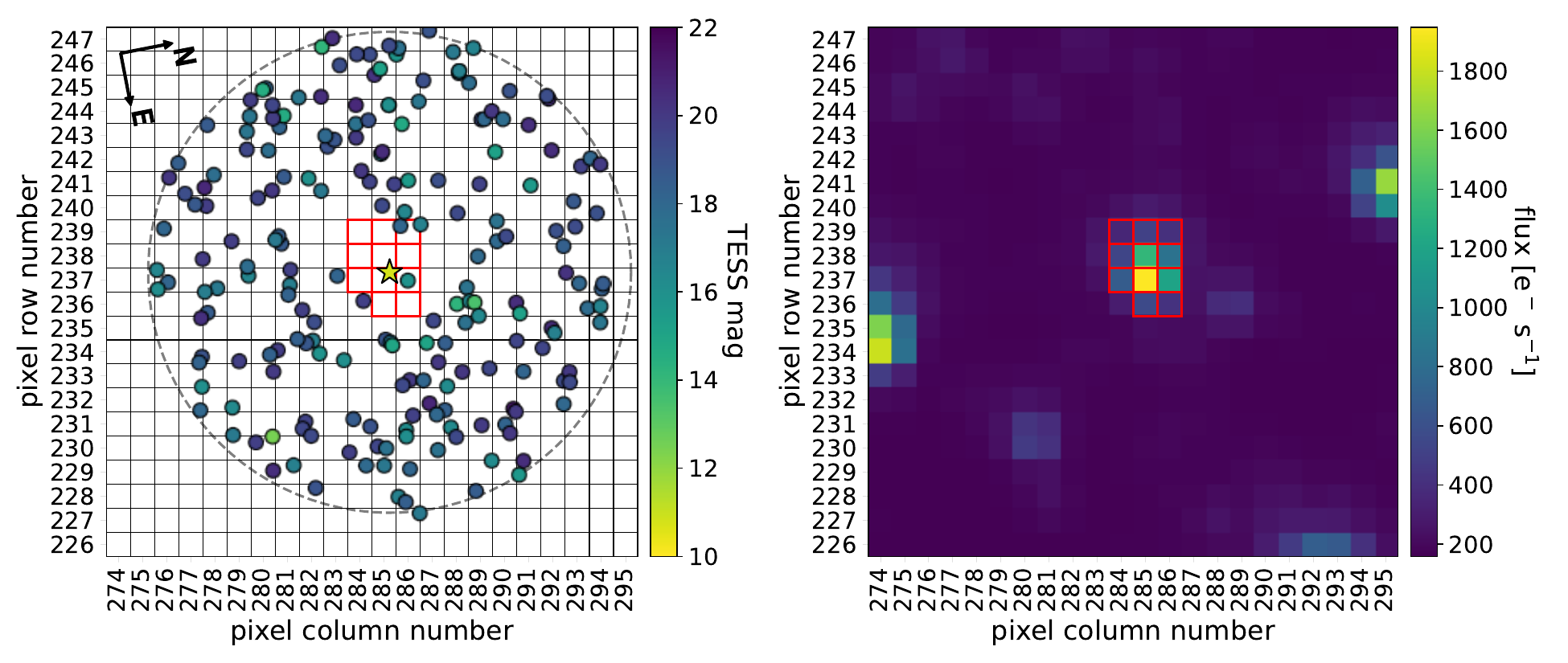}
    \caption{Left panel: the star indicates TOI-791's position on the CCD, and the circles are all sources within 10 pixels (corresponding to approximately $200\arcsec$), coloured according to their \textit{TESS} magnitude. Right panel: sample pipeline aperture overlaid on \textit{TESS} CCD. Figures are for Sector 4 and were produced by {\sc Triceratops}.}
    \label{fig:apertures}  
\end{figure*}

Given TOI-791's position in \tess's southern Continuous Viewing Zone, it was observed \review{in a total of{ 44} sectors: 12} during year 1, \review{13 during year} 3, \review{9 during year} 5, {and 10 during Year 7}. In Fig.~\ref{fig:tess} we present the 2-minute cadence photometry for TOI-791, as well as long cadence photometry for those sectors where 2-minute was not available. In Fig.~\ref{fig:apertures} we present a sample \tess CCD cutout (in this case, Sector 4) overlaid with a typical aperture, as well as all nearby stars.

\subsection{\textit{TESS} Initial Candidate Identification}


All TESS 2-minute cadence data are processed by the Science Processing Operations Center (SPOC) pipeline at NASA Ames Research Center, presented in \cite{SPOC}. Data products are subsequently available to the community in the form of Simple Aperture Photometry \citep[SAP][]{twicken:PA2010SPIE,morris:PA2020KDPH} or Presearch Data Conditioning Simple Aperture Photometry \citep[PDCSAP][]{Stumpe2012,Smith2012,Stumpe2014}, where the latter has been corrected for instrument systematics and crowding effects. Data products can be downloaded from the NASA Mikulski Archive for Space Telescopes (MAST) via the LIGHTKURVE package \citep{lightkurve} or directly through the MAST Portal. All lightcurves are additionally searched by \textsc{SPOC} for periodic transit-like signals and those candidates that meet $\rm SNR\,>\,7.1$ are reported as threshold crossing events (TCEs).

TOI-791.01 was first reported as a CTOI (Community \tess Object of Interest) by Planet Hunters \tess \citep{planethunters} on 2019\,May\,16. The signal was reported with a period of 139.3138~days and a duration of $>12~\rm hours$. It was then adopted as a candidate on 2019\,June\,14 \citep{guerrero2021} following a multi-sector transit search \citep{jenkins2002,jenkins2010,jenkins2020} conducted for sectors 1--9. The transit signal was initially fitted with a limb-darkened transit model \citep{Li2019} and subjected to a suite of diagnostic tests \citep{Twicken2018} to help determine whether the signature is from an exoplanet. The transit signature passed all of the tests reported in the Data Validation report for Sectors 1-9\footnote{All data validation reports \citep{Twicken2018} referred to in this section are downloadable at \url{https://tev.mit.edu/data/search/?q=306472057}}.

\textsc{SPOC} TCEs are subject to the \textsc{TESS-ExoClass} \footnote{\url{https://github.com/christopherburke/TESS-ExoClass}} automated classifier that reduces the number of TCEs that undergo the manual TOI vetting procedure. \textsc{TESS-ExoClass} applies a series of tests that are similar to the \textit{Kepler} \textsc{Robovetter} \citep{coughlin2016,thompson2018}; TOI 791.01 passed all the tests of \textsc{TESS-ExoClass}.

\subsection{Further Candidate Identification}

Given ASTEP's unique capabilities, TOI-791.01 had already been identified as an ideal candidate for our \tess follow-up campaign during the 2020 season. Following the release of Sector 33 at the start of 2021, we identified an unexpected transit-like event at the start of the sector which did not correspond to the known signal. Initial modelling of this event revealed that its duration appeared shorter than that of TOI-791.01, suggesting a shorter orbital period. However, we note that as the transit was exactly at the start of the sector, it was not clear if the ingress was completely observed. The large amount of data available for this CVZ star allowed us to find a possible period at $\sim 110~\rm days$, by assuming that other transits must fall in gaps between sectors. 

The candidate was adopted by ASTEP as TOI-791.02 and we attempted to confirm the period throughout the 2021 and 2022 observing seasons. These observations conclusively ruled out this period. 

\tess resumed observations of the southern celestial hemisphere in January 2023 with Sectors 61-69. A second single transit was detected in Sector 67, some $\sim 928$~days after the Sector 33 event. We modeled both transits individually using {\sc Allesfitter} (see Section \ref{sec:analysis}) and found them to have consistent durations and depths, therefore most likely belonging to the same object. The shortest period alias that was not ruled out by the data is $\sim 232$~days, which would place the planet in a 5:3 period commensurability with TOI-791.01. We adopted a new ephemeris to predict the next event, leading to the observations described in Section \ref{sec:photometry}. \review{We also note that TOI-791.02 was added to ExoFOP as a CTOI by Planet Hunters \textit{TESS} following the Sector 67 event, but as a single transit candidate with no period. Additionally, it was added as a TOI in March 2024 but with incorrect ephemerides attributable to the sparse sampling of the light curve.}

\section{Vetting and Validation}
\label{sec:followup}


In this section, we describe the results of the multi-facility follow-up campaign conducted between November 2019 and March 2024. We begin with the high-resolution imaging observations obtained with Gemini-South and SOAR, and then outline the photometric observations collected from southern observatories in Chile, Australia, South Africa and Antarctica. 

All follow-up observations are summarised in Table \ref{tab:followup}.

\begin{table*}
\centering
\caption{Summary of ground-based follow-up observations carried out for TOI-791.01 and .02}
\begin{tabular}{@{}cccccc@{}}
\midrule \midrule
\multicolumn{6}{c}{\textbf{Follow-up Observations}}                                              \\ \midrule \midrule
\multicolumn{6}{c}{\textbf{High Resolution Imaging}}                                             \\ 
\textbf{Observatory} & \textbf{Filter} & \textbf{Date}     & \textbf{Sensitivity Limit} & \multicolumn{2}{c}{\textbf{Result}}\\ \midrule
Gemini South      & $562~{\rm nm}$     & 2020\,Mar\,12  & $\Delta m=5.1$ at $0.5\arcsec$  & \multicolumn{2}{c}{No sources detected}   \\ 
Gemini South      & $832~{\rm nm}$     & 2020\,Mar\,12  & $\Delta m=6.34$ at $0.5\arcsec$  & \multicolumn{2}{c}{Companion detected}   \\
SOAR      & $879~{\rm nm}$     & 2019\,Nov\,09  & $\Delta m=6.6$ at $1\arcsec$  & \multicolumn{2}{c}{No sources detected}   \\\midrule
\multicolumn{6}{c}{\textbf{Photometric Follow-up}}                                               \\
\textbf{Observatory} & \textbf{Filter} & \textbf{Date}     & \textbf{Coverage} & \textbf{Candidate} &  \textbf{Result} \\ \midrule
El Sauce    & {\it $R_c$}     & 2021\,Feb\,18 & Egress & b  & Detection\\ 
ASTEP400    & {\it $R_c$}     & 2021\,Jul\,06 & Full  & b & Detection\\ 
ASTEP+    & {\it $R_c$} and {\it $V$}     & 2023\,Jun\,03 & Full & b  & Detection\\ 
LCO-SAAO    & {\it $Sloan-i'$}     & 2024\,Feb\,21 & Egress & c & Detection\\ 
Hazelwood    & {\it $R_c$}     & 2024\,Feb\,21 & Egress & c & Detection\\ 
OACC    & {\it $R_c$}     & 2024\,Feb\,21 & Ingress & c & Detection\\ 
LCO-CTIO    & {\it $Sloan-i'$}     & 2024\,Feb\,21 & Out of transit & c & NA\\
LCO-CTIO    & {\it $Sloan-i'$}     & 2024\,Mar\,08 & Ingress & b & Detection\\
Hazelwood    & {\it $R_c$}     & 2024\,Mar\,08 & Egress & b & Detection\\
ASTEP+    & {\it $R_c$} and {\it $V$}     & 2024\,Jul\,25 & Full & b  & Detection\\ 
\review{TRAPPIST-South}   & \review{{\it $R_c$}}    & \review{2025\,Apr\,30} & \review{Ingress} & \review{b}  & \review{Detection}\\ 
\review{ASTEP+ }   &        \review{{\it $R_c$}}    & \review{2025\,Apr\,30} & \review{Egress } & \review{b}  & \review{Detection}\\ 
\review{ASTEP+ }   &        \review{{\it $R_c$} and {\it $V$}} & \review{2025\,May\,30} & \review{Full   } & \review{c}  & \review{Detection}\\
{TRAPPIST-South}   & {{\it $R_c$}}    & {2025\,Sep\,16} & {Ingress} & {b}  & {Detection}\\ 
{ASTEP+ }   &        {{\it $R_c$}}    & {2025\,Sep\,16} & {Egress } & {b}  & {Detection}\\ 
{PEST}   &        {{\it $Sloan-r'$}}    & {2026\,Jan\,17} & {Egress } & {c}  & {Detection}\\\midrule
\multicolumn{6}{c}{\textbf{Spectroscopic Observations}}                                               \\ 
\textbf{Instrument} & \textbf{Wavelength Range} & \textbf{Date}     & \textbf{Number of Spectra} & \multicolumn{2}{c}{\textbf{Use}}\\ \midrule
CORALIE/Euler 1.2m     & $387-689~\rm nm$      &  2019\,Nov - 2022\,Feb & 15  & \multicolumn{2}{c}{Stellar characterisation and mass limits}  \\
FEROS/MPG 2.2m     & $360-920~\rm nm$      &  2019\,Jun - 2020\,Mar & 3  & \multicolumn{2}{c}{Stellar characterisation and mass limits} \\\midrule

\end{tabular}
\label{tab:followup}
\end{table*}

\subsection{High resolution imaging}
\label{sec:hires}

If a star hosting a planet candidate has a close bound or line-of-sight companion, the star will be blended in the \tess light curve and essentially all other measurements of the host star. The companion can then create a false-positive exoplanet detection if it is an eclipsing binary (EB) or other variable. Flux from this additional source can lead to an underestimated planetary radius if not accounted for in the transit model \citep{Ciardi2015}, incorrect derived exoplanet and stellar properties \citep{FH2017, FH2020}, and complete dilution of additional small planet transits within the same system, making them undetectable \citep{Lester2021}. The presence of a close companion star does not automatically invalidate the measured transits and their interpretation as signals of exoplanetary origin but provides knowledge allowing proper correction of their ``third-light” contribution. In order to search for close companions, TOI-791 was subjected to high-resolution optical imaging using the technique of speckle interferometry.

\subsubsection{Gemini-South}

\begin{figure}
    \centering
    \includegraphics[width=\columnwidth]{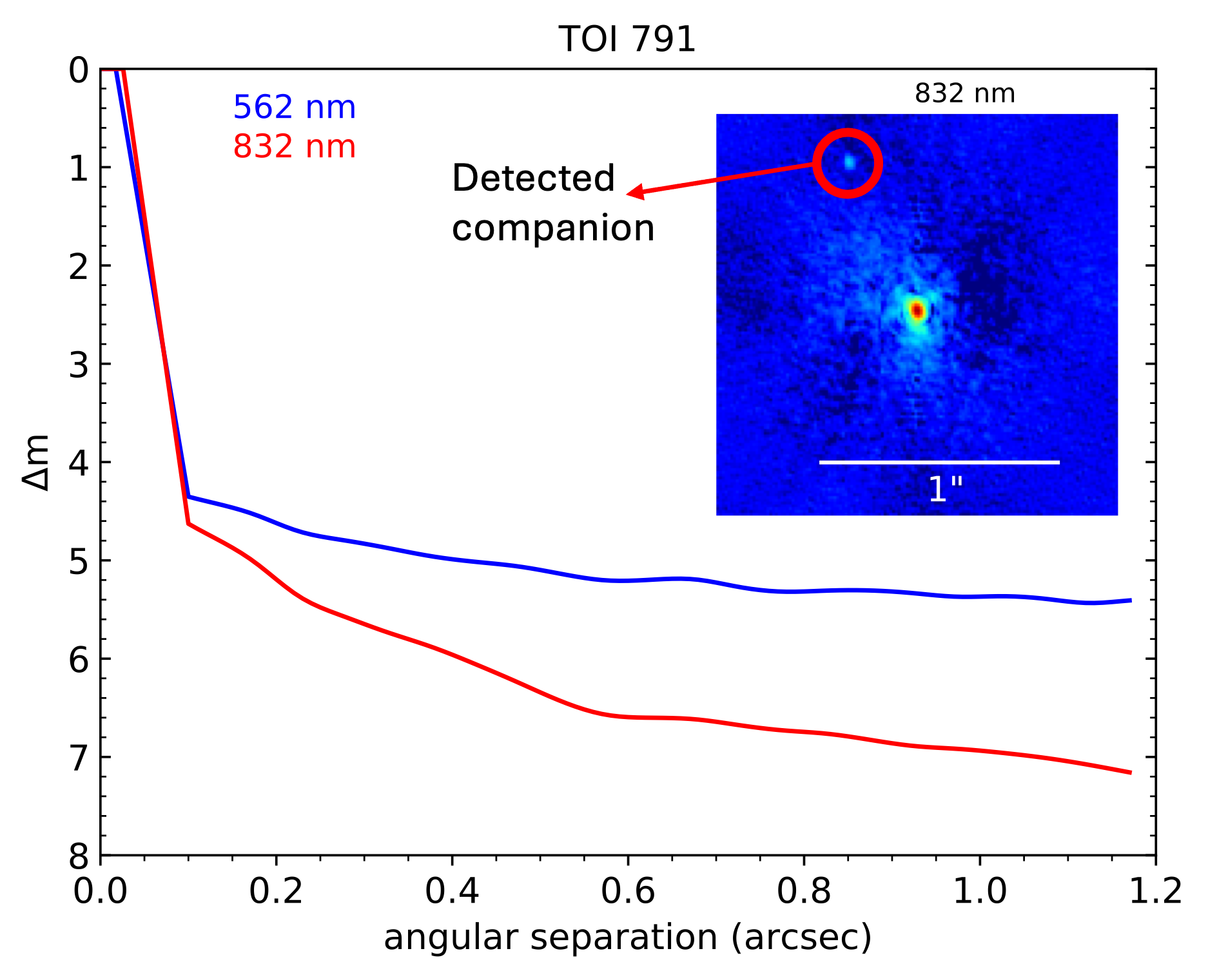}
    \caption{Plot showing the $5\sigma$ speckle imaging contrast curves in both filters as a function of the angular separation out to $1.2\arcsec$, the end of speckle coherence \review{obtained using Zorro on Gemini-South}. The inset shows the reconstructed $832~\rm nm$ image with a $1\arcsec$ scale bar. The star, TOI-791, was found to have a close companion at $0\farcs709$ which can be seen in the reconstructed image.}
    \label{fig:gemini}
\end{figure}

TOI-791 was observed on 2020 March 12 UT using the Zorro speckle instrument on the Gemini South 8-m telescope \citep{Scott2021}. Zorro provides simultaneous speckle imaging in two bands (562 nm and 832 nm) with output data products including a reconstructed image with robust contrast limits on companion detections \citep[e.g.][]{Howell2016}. Five sets of 1000 exposures of 0.06~s were collected and subjected to Fourier analysis in our standard reduction pipeline \citep{Howell2011}. Fig.~\ref{fig:gemini} shows our achieved 5$\sigma$ magnitude contrast curves and the 832~nm reconstructed speckle image. 

We find that TOI-791 has a close companion star as seen in the 832~nm reconstructed image. The companion star was only detected in the 832~nm observation and has a measured separation of 0.709$\arcsec$, lies at a position angle of 25.648~degrees, and has a delta magnitude (at 832~nm) of 5.6 magnitudes. Taking the Galactic location of TOI-791 (latitude = -20, longitude = 282 degrees) and the work presented in \citep{Matson2018} giving probabilities for close companions being bound or line-of-sight, this companion star has a probability of 85\% of being gravitationally bound. The large magnitude difference between the primary star and the companion and its detection only at the red wavelength suggest that the companion is approximately an M1V star with a mass near $0.45 {\rm M}_{\odot}$. Within the angular limits of the speckle observations (8-m diffraction limit out to $1.2\arcsec$) and the 5$\sigma$ contrast limits, no other close companion was detected.  At the distance of TOI-791 ($d=341$~pc) these angular limits correspond to spatial limits of $6.8$ to $409$~au.

\subsubsection{SOAR}

\begin{figure}
    \centering
    \includegraphics[width=\columnwidth]{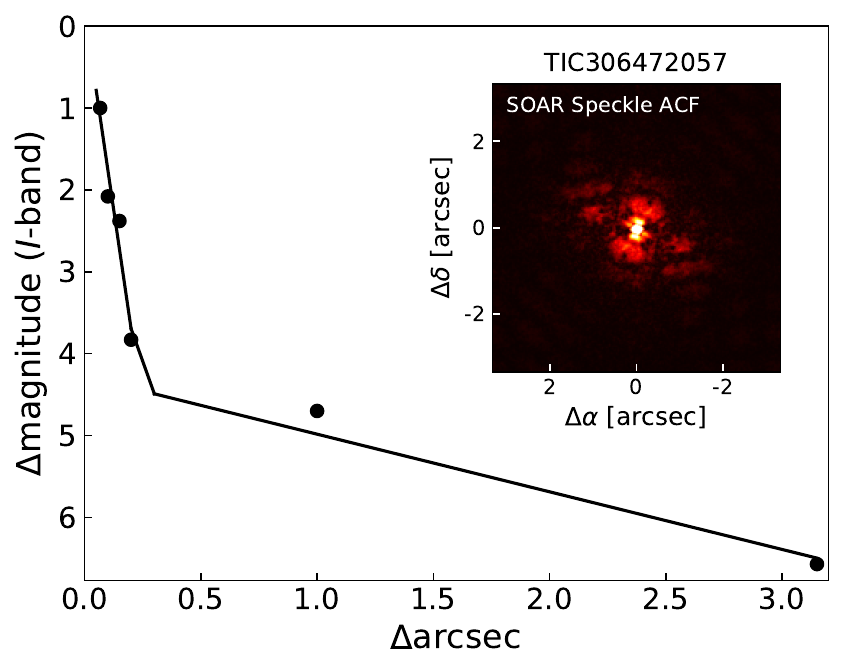}
    \caption{Plot showing the $5\sigma$ speckle imaging contrast curves obtained with Zorro in \textit{Cousins I}-Band as a function of the angular separation out to $3\arcsec$. The inset shows the reconstructed speckle image; no companions were detected in the SOAR imaging.}
    \label{fig:soar}
\end{figure}

High-angular resolution imaging is needed to search for nearby sources that can contaminate the \tess photometry, resulting in an underestimated planetary radius, or be the source of astrophysical false positives, such as background eclipsing binaries. We searched for stellar companions to TOI-791 with speckle imaging on the 4.1-m Southern Astrophysical Research (SOAR) telescope \citep{soar} on 9 November 2019 UT, observing in \textit{Cousins I}-band, a similar visible bandpass as \tess. This observation was sensitive with 5-sigma detection to a 4.9-magnitude fainter star at an angular distance of 1 arcsec from the target. More details of the observations within the SOAR \tess survey are available in \cite{soar2}. The 5$\sigma$ detection sensitivity and speckle auto-correlation functions from the observations are shown in Fig.~\ref{fig:soar}. No nearby stars were firmly detected within 3\arcsec of TOI-791 in the SOAR observations.

\subsection{Photometric follow-up}
\label{sec:photometry}

\subsubsection{ASTEP}


ASTEP (the Antarctic Search for Transiting ExoPlanets) is a $0.4{\rm m}$ Newtonian telescope hosted at Concordia Station on the Antarctic Plateau \citep{Guillot2015}. Until the end of the 2021 observing season, ASTEP was equipped with a 4k $\times$ 4k front-illuminated FLI Proline KAF-16801E CCD with an image scale of $0\farcs93\,{\rm px}^{-1}$ resulting in a $1^{\circ} \times 1^{\circ}$ corrected field of view. The focal instrument dichroic plate split the beam into a blue wavelength channel for guiding, and a non-filtered red science channel roughly matching an {\it Rc} transmission curve \citep{Abe2013}.

As of the start of the 2022 observing season, the focal box was replaced with a new one containing two 2k $\times$ 2k backside illuminated cameras (ANDOR iKon-L 936 and FLI Kepler KL400) allowing simultaneous observations in red and blue channels respectively, with transmission curves centred on $850~{\rm nm}$ and $555~{\rm nm}$ \citep{astep+}. 

Due to the extremely low data transmission rate from Concordia Station to the rest of the world, the data are processed automatically on-site using {\sc IDL} \citep{idlpipeline} and {\sc Python} \citep{ASTEP, prosepaper, prosesoft} pipelines. 

We observed three full transits of TOI-791.01 on UTC 2021 July 06, UTC 2023 June 03, and UTC 2024 July 25 \review{and partial transits on UTC 2025 April 30} {and UTC\,2025\,September\,16}. The first of these was observed before the new camera box was installed, and thus the data was collected only in the red channel. For the \review{three} most recent observations, however, we collected data simultaneously in both the red and blue channels, \review{although in the most recent observations the data from the blue camera were of very poor quality and the events were not recoverable}. 

In addition, attempts to detect the transit of TOI-791.02 were performed on 2021-03-28, 2022-05-04 and 2024-05-08, all leading to flat lightcurves, ruling out the ephemerides that had been previously calculated. \review{A full transit was observed and detected on UTC 2025 May 30 simultaneously in both the red and blue channels.} {This event confirmed the $\sim 232$~day period for TOI-791.02 as it becomes the longest possible alias. All shorted period aliases are ruled out by the \textit{TESS} data.}

\subsubsection{El Sauce}


We observed an egress in Johnson-Cousins $R_c$-band on UTC 2021 February 18 using the Evans 0.36m telescope at El Sauce Observatory in Coquimbo Province, Chile. The telescope was equipped with a ST1603-3 CCD camera with $1536 \times 1024$ pixels binnedu $2 \times 2$ in-camera resulting in an image scale of $1.47\arcsec/{\rm px}$. The photometric data was obtained from $516 \times 60$  seconds exposures, after standard calibration, using a circular $7.4\arcsec$ aperture in {\sc AstroImageJ} \citep{Collins:2017}.

\subsubsection{Hazelwood}
We observed an egress of TOI-791.02 and of TOI-791.01, on UTC 2024 February 21 and UTC 2024 March 8, respectively in Rc-band  from Hazelwood Observatory near Churchill, Victoria, Australia., The f/8 0.32\,m telescope is equipped with a $2184\times1472$ SBIG STT3200 camera. The image scale is $0\farcs56$ per pixel, resulting in a $20\arcmin\times14\arcmin$ field of view. Differential photometric data were extracted using {\sc AstroImageJ} \citep{Collins:2017} using a circular $5\farcs0$ photometric aperture.

\subsubsection{LCOGT}
We observed a partial transit and an out-of-transit window of TOI-791.01 and TOI-791.02, respectively, on UTC 2024 February 21 in Sloan $i'$ from the Las Cumbres Observatory Global Telescope (LCOGT) \citep{Brown:2013} 1\,m network node at Cerro Tololo Inter-American Observatory in Chile (CTIO). We observed another partial transit window of TOI-791.02 on UTC 2024 February 21 in Sloan $i'$ from the LCOGT 1\,m network node at South Africa Astronomical Observatory near Sutherland, South Africa (SAAO). We used the {\tt TESS Transit Finder}, which is a customized version of the {\sc Tapir} software package \citep{Jensen:2013}, to schedule our transit observations. The 1\,m telescopes are equipped with a $4096\times4096$ SINISTRO camera having an image scale of $0\farcs389$ per pixel, resulting in a $26\arcmin\times26\arcmin$ field of view. All images were calibrated by the standard LCOGT {\sc BANZAI} pipeline \citep{McCully:2018}, and differential photometric data were extracted using {\sc AstroImageJ} \citep{Collins:2017}.  We used circular photometric apertures with radii ranging from $5\farcs4$--$6\farcs2$ that excluded all of the flux from the nearest known neighbor in the \gaia DR3 catalog (\gaia DR3 5270467966514084864) that is $17\farcs$ northwest of TOI-791. 


\subsubsection{OACC-CAO}
We observed with the Campocatino Austral Observatory (CAO), located in El Sauce, Chile. The CAO is a PlaneWave CDK24 Corrected Dall-Kirkham telescope on a Planewave L350 equatorial mount, equipped with a Moravian C3-61000E CMOS camera. We observed in \textit{sloan-i'} filter for $\approx$ 8 hours on UTC 2024\,February\,21. We used a $14\arcsec$ photometric aperture and reached a photometric precision of $0.7~\rm ppt$ after averaging our data points over a time-scale of $28\rm ~minutes$.

\subsubsection{\review{TRAPPIST-South}}

\review{We used the 0.6m-TRAPPIST-South \citep[TRAnsiting Planets and PlanetesImals Small Telescope,][]{Jehin2011,Gillon2011} to observe an ingress on UTC\,2025\,April\,29 {and UTC\,2025\,September\,16}.
The observations were carried out in the R filter with an exposure time of 20s. The telescope is equipped with a 2K$\times$2K FLI Proline CCD camera with a FOV of $22\arcmin\times22\arcmin$ and a pixel scale of 0.64\arcsec/pixel. The science images calibration and photometric extraction were performed using the {\sc PROSE} pipeline \citep{prosesoft,prosepaper}.}

\subsubsection{{PEST}}

{We observed an egress of TOI-791.02 on UTC 2026 January 17 in Sloan r' using the CDK14 telescope at the Perth Exoplanet Survey Telescope (PEST) observatory located near Perth, Australia.  The f/7.2 0.365 m telescope is equipped with a $6252\times4176$ QHY268M camera. Images are binned 2x2 in software giving an image scale of 0$\farcs$6 pixel$^{-1}$ resulting in a $32\arcmin\times21\arcmin$ field of view. A custom pipeline based on {\tt C-Munipack}\footnote{http://c-munipack.sourceforge.net} was used to calibrate the images and extract differential photometry using a circular $3\farcs0$ photometric aperture.}

\section{Global photometric analysis}
\label{sec:analysis}


We use \textsc{Allesfitter} \citep{AllesfitterPaper,AllesfitterSoft} to jointly model the full photometric dataset described in Section \ref{sec:followup} together with the \tess photometry. \textsc{Allesfitter} is a flexible inference package for exoplanet modelling written in \textsc{Python}. It uses {\sc Ellc} to generate lightcurve models \citep{ellc} and \textsc{Celerite} to generate Gaussian Process (GP) models \citep{celerite}. Best fitting models are chosen using MCMC sampling via \textsc{emcee} \citep{emcee}, or nested sampling via \textsc{Dynesty} \citep{dynesty}. 

We adopt the transit parameters from Section \ref{sec:sherlock} as uniform priors, and the stellar parameters described in Section \ref{sec:star} as normal priors, and fit for all planetary parameters ($R_{\rm p}/R_\star$, $(R_\star+R_{\rm p})/a$, $\cos\,i$, $T_0$, and $P$). Additionally, we make use of \textsc{PyLDTK} \citep{pyldtk} and Pheonix stellar atmosphere models \citep{phoenix} to calculate quadratic limb darkening coefficients, which we reparametrise in the \cite{kippingldcs} parametrisation and adopt as normal priors in our fit. 

We use the nested sampling algorithm in all our fits; we find this the best choice for modelling this system as the Bayesian evidence is calculated at each step, allowing us to compare the evidence and select the most statistically favoured model. 

One peculiarity of this system is that the outer planet has shorter transits than the inner planet. This can be explained by a small mutual inclination, eccentric orbits, an oblate star, or a combination of these factors (as is likely). To find the best-fitting solution, we consider two models in the first instance: one with free eccentricity and one where the orbits are constrained to be circular. In both cases eccentricity is parametrised as $\sqrt{e_{\rm b}} \cos{\omega_{\rm b}}$ and $\sqrt{e_{\rm b}} \sin{\omega_{\rm b}}$ {\citep[as in][]{Triaud2011}}; we also impose a wide uniform prior on $\cos{i}$ to allow for mutual inclinations. 

Baseline trends can be modelled in various ways through \textsc{Allesfitter}, the simplest being the use of a `hybrid spline' and in most cases this is sufficient. However, we instead opt for a GP for three of our lightcurves. The CCAO lightcurve was affected by poor weather in the second half of the data which can be modelled out with a GP. The ASTEP lightcurve of 2021 also has a trend that we can remove with a GP caused by a brief twilight which interrupted the transit, and caused the sky to brighten temporarily during the observation. Finally, we also use a GP to detrend the \textit{TESS} data. The `hybrid spline' performs best on continuous data sets, but to speed up processing we only model the \textit{TESS} in two-day wide windows centred on each transit. The GP is therefore the best approach to remove any trends that would alter the transit depths. We choose the 3/2 Mat\'{e}rn kernel due to its flexibility and its usefulness in removing stochastic trends from our lightcurves.



The \textit{TESS} PDCSAP lightcurves are already corrected for contamination from nearby sources, however, our high resolution imaging detailed in Section \ref{sec:hires} revealed the presence of a blended M dwarf companion. We therefore calculate the flux contribution from this additional contaminating source and calculate a dilution for \tess and ground-based lightcurves in red channels of $0.00572$. As the companion was not detected in the blue high resolution imaging, we set the dilution in the blue channel observations to zero. We fix the dilution coefficients in all our fits. 

The results of our fits indicate that the data are consistent with eccentric orbits, with the free eccentricity model preferred over the circular model with a Bayes Factor (BF) $\sim 124$. However, when eccentricity is left completely free, the model finds very high eccentricities of $e_b=0.49$ and $e_c=0.43$. The result of these highly eccentric orbits is that they cross. While this is not impossible, especially in cases of objects in a mean motion resonance \citep[e.g. Neptune and Pluto][]{forgacs2018}, it is not likely; it is also unprecedented for an exoplanetary system. We therefore run three additional fits of the data attempting to limit the eccentricities by narrowing the uniform priors on $\sqrt{e_{\rm b}} \cos{\omega_{\rm b}}$ and $\sqrt{e_{\rm b}} \sin{\omega_{\rm b}}$ to [$-0.25, 25$], [$-0.5, 5$], and [$-0.75, 75$]. We find that the preferred solution comes from limiting the prior to [$-0.5, 0.5$]; this solution still yields eccentric orbits but they do not cross. This solution is favoured over the free model with a BF$\sim 10$ and over the next best solution ([$-0.25, 25$]) with a BF$\sim 5$.
While non-crossing orbits solutions are consistent with the photometric data, their orbital spacing make the configuration unstable even at low eccentricities if the planet masses are too high.
In our dynamical analysis (see section \ref{sec:nepdes}), we found that even at low eccentricities the best-fit solutions were unstable on short timescales.
Additionally, because of the proximity to the 5:3 mean motion resonance, moderate eccentricities would lead to transit timing variations of several hours unless the planets are trapped deep into the resonance.
As a result, these dynamical considerations make us prefer the circular fit until proper photodynamical modeling can be done for the system.
We adopt this solution and present all fitted and derived parameters in Table~\ref{tab:glob_fit}.

{The result of fixing the eccentricity to zero is that the shorter transit duration of the outer planet is explained by the difference in impact parameter. As the outer planet has a higher impact parameter, it crosses a shorter chord across the stellar disc.}

\begin{figure}
    \centering
    \includegraphics[width=\columnwidth]{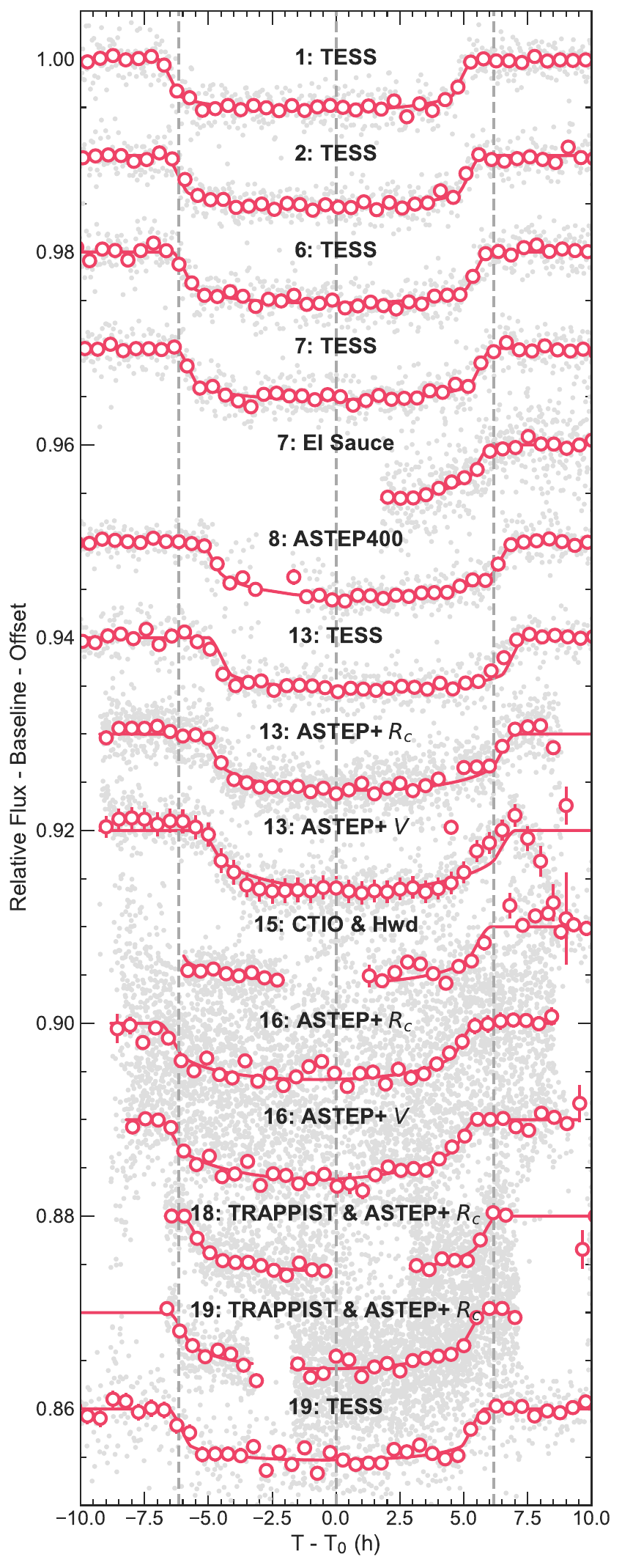}
    \caption{Individual transit lightcurves for TOI-791~b. The grey points are the raw flux while the pink circles are binned to 30 minutes. The vertical grey dashed line\review{s} represents the linear ephemeris prediction \review{for ingress, midtime and egress. The number refers to the transit epoch number, with `1' being the first transit observed. We note that some transit epochs were observed simultaneously with multiple instruments. Transits are offset by $0.01 \times (\rm transit number - 1).$}}
    \label{fig:planet_b}
\end{figure}

\begin{figure}
    \centering
    \includegraphics[width=\columnwidth]{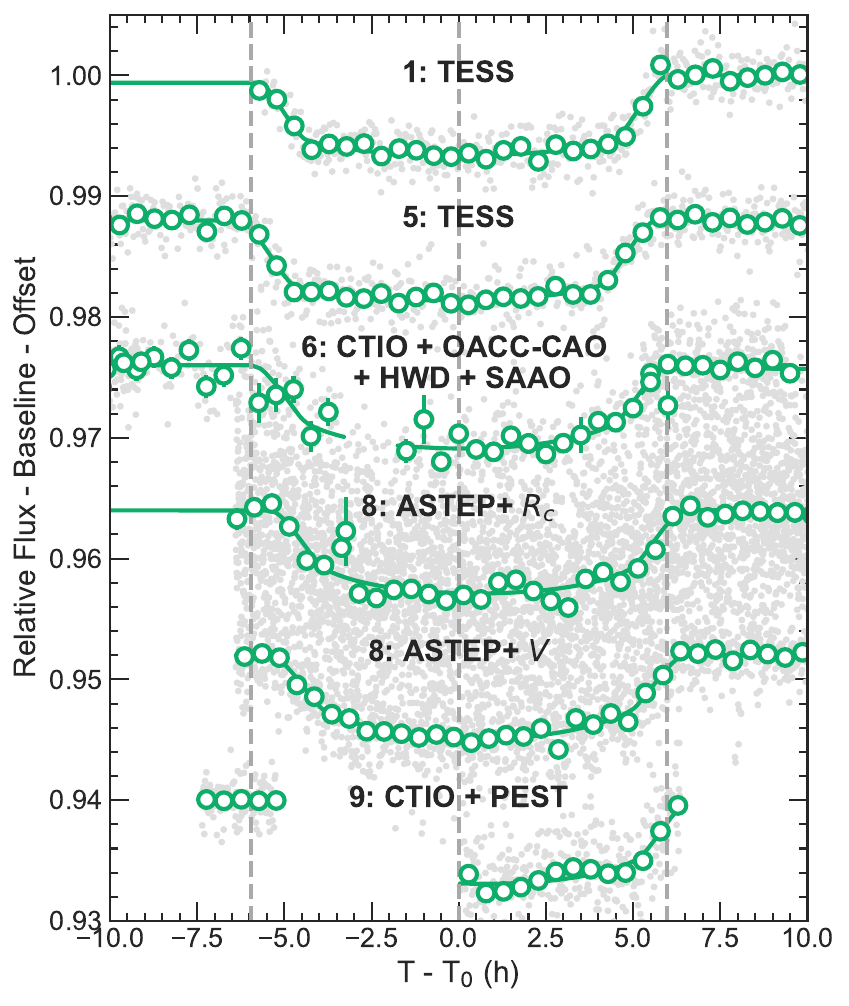}
    \caption{Individual transit lightcurves for TOI-791~c. The grey points are the raw flux while the green circles are binned to 30 minutes. The vertical grey dashed line\review{s} represents the linear ephemeris prediction \review{for ingress, midtime and egress. The number refers to the transit epoch number, with `1' being the first transit observed. We note that some transit epochs were observed simultaneously with multiple instruments. Transits are offset by $0.012
    \times (\rm transit number - 1).$}}
    \label{fig:planet_c}
\end{figure}

In Figures \ref{fig:planet_b} and \ref{fig:planet_c} we present all individual transits collected for planets b and c respectively.

\begin{table*}
\centering
\caption{Priors used in our fit, along with fitted and derived parameters. Uniform priors are indicated as $\rm \mathcal{U}(lower~bound,~upper~bound)$ and normal priors are indicated as $\rm \mathcal{N}(mean,~standard~deviation)$. $^\star$Equilibrium temperature is calculated assuming an albedo of 0.3 and emissivity of 1.}
\resizebox{\textwidth}{!}{%
\begin{tabular}{@{}cccc@{}}
\toprule
\multicolumn{2}{c}{\textbf{External Priors}}  & \multicolumn{2}{c}{\textbf{GP Priors}}                              \\\midrule
Stellar Mass; $M_{\star}$ ($\mathrm{M_{\odot}}$)  & $\mathcal{N}(1.28, 0.08)$ & Amplitude Scale $\mathrm{GP \ln \sigma (flux)}$ & $\mathcal{U}(-7, -7)$      \\
Stellar Radius;  $R_{\star}$ ($\mathrm{R_{\odot}}$)  & $\mathcal{N}(1.474, 0.0047)$  & Length scale $\mathrm{GP \ln \rho (flux)}$   & $\mathcal{U}(-7, 2)$  \\
Stellar Effective Temperature; $T_{\rm eff}$ (K) & $\mathcal{N}(6294, 128)$ &                     &              \\\midrule
\multicolumn{4}{c}{\textbf{Limb Darkening Coefficients}} \\ \midrule
\tess $\rm u_1$ & $0.280_{-0.040}^{+0.036}$ & \tess $\rm q_1$ & $\mathcal{N}(0.359, 0.050)$ \\
\tess $\rm u_2$ & $0.214_{-0.039}^{+0.055}$ & \tess $\rm q_2$ & $\mathcal{N}(0.338, 0.050)$  \\
{\it $Sloan-i'$} $\rm u_1$ & $0.441_{-0.051}^{+0.058}$ & \textit{R$_{\rm c}$} $\rm q_1$ & $\mathcal{N}(0.332, 0.050)$ \\
{\it $Sloan-i'$} $\rm u_2$ & $0.171\pm0.047$ & \textit{R$_{\rm c}$} $\rm q_2$ & $\mathcal{N}(0.356, 0.050)$  \\ 
\textit{R$_{\rm c}$} $\rm u_1$ & $0.537_{-0.044}^{+0.055}$ & \textit{R$_{\rm c}$} $\rm q_1$ & $\mathcal{N}(0.478, 0.050)$ \\
\textit{R$_{\rm c}$} $\rm u_2$ & $0.198_{-0.051}^{+0.043}$ & \textit{R$_{\rm c}$} $\rm q_2$ & $\mathcal{N}(0.348, 0.050)$  \\ 
\textit{V} $\rm u_1$ & $0.781_{-0.061}^{+0.070}$ & \textit{V} $\rm q_1$ & $\mathcal{N}(0.503, 0.050)$ \\
\textit{V} $\rm u_2$ & $-0.019_{-0.073}^{+0.068}$ & \textit{V} $\rm q_2$ & $\mathcal{N}(0.383, 0.050)$  \\\midrule
\multicolumn{4}{c}{\textbf{Dilution}}                              \\\midrule
\tess  & $0.00572$ (fixed) & Ground-based \textit{$R_c$} & $0.00572$ (fixed)  \\ \midrule 
\multicolumn{3}{c}{\bf Fitted GP Hyperparameters} &  \textcolor{black}{\textbf{Source}}  \\ \midrule
\tess  & $\mathrm{gp \ln \sigma}$  $-7.42\pm0.12$ & $\mathrm{gp \ln \rho }$ $-1.19_{-0.24}^{+0.23}$  & linear circular fit \\
ASTEP400  & $\mathrm{gp \ln \sigma}$  $-6.39_{-0.18}^{+0.20}$ & $-2.64_{-0.16}^{+0.22}$  & linear circular fit \\ 
CAAO  & $\mathrm{gp \ln \sigma}$  $-4.87_{-0.36}^{+0.45}$ & $\mathrm{gp \ln \rho }$ $-2.25_{-0.36}^{+0.41}$  & linear circular fit \\ 
ASTEP+ (\textit{$R_c$})  & $\mathrm{gp \ln \sigma}$  $-5.98_{-0.24}^{+0.31}$ & $\mathrm{gp \ln \rho }$ $-2.44\pm0.30$  & linear circular fit \\ 
ASTEP+ (\textit{V})  & $\mathrm{gp \ln \sigma}$  $-5.66_{-0.25}^{+0.29}$ & $\mathrm{gp \ln \rho }$ $-2.84_{-0.28}^{+0.35}$  & linear circular fit \\ \midrule
{ \bf }     & { \bf TOI-791~b}  & { \bf TOI-791~c}    &  \textbf{Source}      \\\midrule 
\multicolumn{4}{c}{\textbf{Fit Parametrisation and Priors}}                           \\ \midrule
Transit Depth; $ R_{\rm p} / R_\star$         & $\mathcal{U}(0.05, 0.1)$   & $\mathcal{U}(0.05, 0.1)$   \\
Inverse Scaled Semi-major Axis; $(R_\star + R_{\rm p}) / a$   & $\mathcal{U}(0.008, 0.015)$    & $\mathcal{U}(0.005, 0.013)$    \\
Orbital Inclination; $\cos{i}$               & $\mathcal{U}(0.000, 0.01)$    & $\mathcal{U}(0.000, 0.01)$   \\
Transit Epoch; $T_{0}$ (BJD)  & $\mathcal{U}(2\,459\,263.35, 2\,459\,263.55)$   & $\mathcal{U}(2\,459\,433.9, 2\,459\,434.1)$  \\
Period; $P$  (days)          & $\mathcal{U}(139.2, 139.4)$    & $\mathcal{U}(231.9, 232.1)$   & \\
\midrule
\multicolumn{4}{c}{{ \bf Fitted Parameters}}                    \\ \midrule
$R_{\rm p} / R_\star $      & \textbf{$0.06924\pm0.00054$}  &    \textbf{$0.0805\pm0.0011$}    & fixed linear ephem. fit     \\
$(R_\star + R_{\rm p}) / a$         & \textbf{$0.01215_{-0.00021}^{+0.00025}$} & \textbf{$0.00858\pm0.00017$}   & fixed linear ephem. fit \\
$\cos{i}$       & \textbf{$0.00355_{-0.00063}^{+0.00068}$ }  & \textbf{$0.00532\pm0.00023$}    &  fixed linear ephem. fit    \\
$T_{0}$ $(\mathrm{BJD})$    & \textbf{$2459263.45083\pm0.00075$} & \textbf{$2459433.98252\pm0.0015$ }  & linear circular fit \\
$P$ $(\mathrm{d})$       & \textbf{$139.29931_{-0.00012}^{+0.00011}$} & \textbf{$232.01570_{-0.00071}^{+0.00067}$}       & linear circular fit  \\ 
\midrule
\multicolumn{4}{c}{{ \bf Derived Parameters}}        \\ \midrule
Companion radius; $R_\mathrm{p}$ ($\mathrm{R_{jup}}$)          & \textbf{$0.993\pm0.033$}    & \textbf{$1.155\pm0.040$}                  & fixed linear ephem. fit  \\
Semi-major axis; $a$ (AU)                                      & \textbf{$0.602\pm0.022$}    & \textbf{$0.863\pm0.032$}                 & fixed linear ephem. fit \\
Inclination; $i$ (deg)                                         & \textbf{$89.797_{-0.039}^{+0.037}$ }  & \textbf{$89.695\pm0.013$}   & fixed linear ephem. fit \\
Impact parameter; $b$                                          & \textbf{$0.313\pm0.053$}    & \textbf{$0.671\pm0.017$}         & fixed linear ephem. fit  \\
Total transit duration; $T_\mathrm{tot}$ (h)                   & \textbf{$12.366\pm0.050$}   &  \textbf{$11.926\pm0.082$}         & fixed linear ephem. fit  \\
Full-transit duration; $T_\mathrm{full}$ (h)                   & \textbf{$10.599_{-0.048}^{+0.040}$}   &  \textbf{$8.853\pm0.097$ }   & fixed linear ephem. fit  \\
Equilibrium temperature$^{\star}$; $T_\mathrm{eq}$ (K)         & \textbf{$434.3\pm9.7$}    & \textbf{$362.8\pm8.2$}                     & fixed linear ephem. fit  \\
Instellation; $S$ ($S_{\oplus}$)                               & \textbf{$8.47\pm2.06$ }            & \textbf{$4.12\pm1.01$ }                  & fixed linear ephem. fit  \\
Transit depth \tess; $\delta_\mathrm{tr; TESS}$ (ppt)          &  \textbf{$5.428\pm0.078$}    & \textbf{$6.78\pm0.17$}          & fixed linear ephem. fit      \\
Transit depth \tess; $\delta_\mathrm{tr; Sloan-i'}$ (ppt)      &  \textbf{$5.669_{-0.075}^{+0.082}$ }   & \textbf{$6.85\pm0.17$}  & fixed linear ephem. fit      \\
Transit depth \textit{R$_c$}; $\delta_\mathrm{tr; R_C}$ (ppt)  &  \textbf{$5.864\pm0.083$}        & \textbf{$6.93_{-0.17}^{+0.18}$}       & fixed linear ephem. fit  \\
Transit depth \textit{V}; $\delta_\mathrm{tr; V}$ (ppt)        &  \textbf{$6.186_{-0.093}^{+0.10}$} & \textbf{$6.93\pm0.19$}      & fixed linear ephem. fit      \\ \midrule

\end{tabular}%
}
\label{tab:glob_fit}
\end{table*}

\subsection{Transit timing variations}

In Section \ref{sec:sherlock} we indicated that the two planets orbiting TOI-791 find themselves very close to a 5:3 mean-motion resonance. We therefore run an additional fit of the data with a fixed linear ephemeris to check for the presence of Transit Timing Variations (TTVs). 

To speed up this fit, in addition to fixing the linear ephemeris we also fix all limb-darkening coefficients and GP parameters to those extracted from the circular linear fit. This means we are only fitting the individual transit midtimes, the error scaling on the flux and system parameters $R_{\rm p}/R_\star$, $(R_\star+R_{\rm p})/a$, and $\cos\,i$. 

\begin{figure}
    \centering
    \includegraphics[width=\columnwidth]{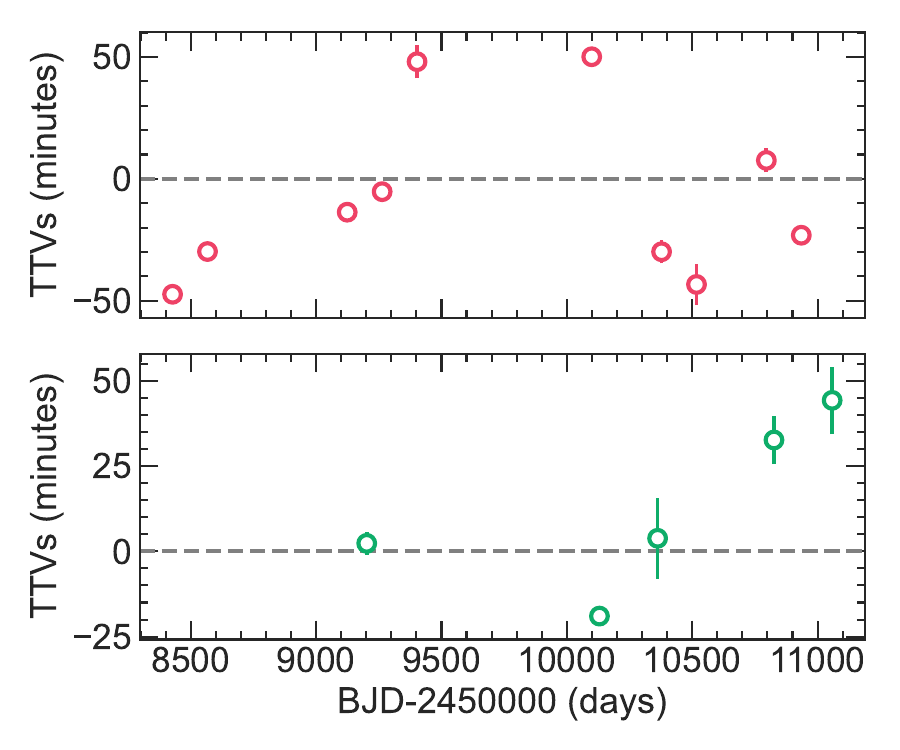}
    \caption{Transit timing variations for TOI-791~b (upper panel) and c (lower panel).}
    \label{fig:ttvs}
\end{figure}

The results of these fits are presented in Figure \ref{fig:ttvs}. For planet b, where we have observed a total of \review{nine} transits, we measure TTVs of up to $50~{\rm min}$ over the whole baseline of observations, \review{while for planet c we measure TTVs $>35~\rm min$ for the four transits observed. } 

The variation in TTVs over time for planet b appear sinusoidal, however given that the `super-period' \citep{Lithwick2012} of this system is expected to be in excess of 88 years, the variation we see is in fact the short-term `chopping' signal. Significantly more transits will be needed to reveal the long-term TTV signal. 

\section{Constraints on the planet masses}
\label{sec:mass_limits}

\subsection{\review{Radial velocity constraints}}

We analyse the radial velocity measurements from the CORALIE and FEROS spectra to obtain upper limits on the masses of the two bodies confirming that they are in the planetary regime. We perform the analysis using {\sc kima} \citep{kima}. \review{We use a two planet model by fixing the number of free planets to 0, and using {\sc kima}'s {\it known object} mode, allowing for different priors on two Keplerians. {\sc kima} fits Keplerian orbits using diffusive nested sampling \citep[{\sc DNEST4};][]{brewer_dnest4_2018}. The likelihood evaluation is performed using a Student's t distribution, this allows for for outliers to be accounted for by having a lesser effect on the likelihood due to the wide tails of the student's t distribution. A jitter term is also included to account for astrophysical noise in the data. When analysing the two datasets, an offset is also included to account for different instrument systematics between CORALIE and FEROS. {\sc kima} has a two-step process. First likelihood levels are built until a convergence criterion has been acheived, then a sampling stage occurs for a pre-determined number of steps. Here we use 200\,000 steps in each analysis\footnote{These each ran for \(\sim 2\) hrs on 4 cores of a Intel® Xeon® Platinum 8360Y}.}

We fix the number of planets in the fit to \review{two} and enforce priors based on the photometric fit. We perform the analysis twice with two different sets of eccentricities: one based on the best fitting photometric model when eccentricity is left free, and one where approximately circular orbits are assumed. \review{Priors used in each analysis are shown in Table \ref{tab:kima_priors}.}

\begin{table}
    \centering
    \caption{\review{Prior distributions used in the {\sc kima} analysis of radial velocities. Shown for both of the two analyses restricting to circular or eccentric orbits. Distributions are Normal \(\mathcal{N}\), Uniform \(\mathcal{U}\), Log-Uniform \(\mathcal{LU}\), and Modified Log-Uniform \(\mathcal{MLU}\) (which is Uniform between 0 and the lower value, then Log-Uniform between the two values).}}
    \begin{tabular}{ccr}
         Parameter & Priors, Circular orbits & Priors, Eccentric orbits \\
         \hline
         \(P_{b}\) [days] & \(\mathcal{N}(139.3,0.1)\) & \(\mathcal{N}(139.3,0.1)\) \\
         \(P_{c}\) [days] & \(\mathcal{N}(232.0,0.1)\) & \(\mathcal{N}(232.0,0.1)\) \\
         \(K_{b}\,[{\rm m\,s^{-1}}]\) & \(\mathcal{LU}(1,10000)\) & \(\mathcal{LU}(1,10000)\) \\
         \(K_{c}\,[{\rm m\,s^{-1}}]\) & \(\mathcal{LU}(1,10000)\) & \(\mathcal{LU}(1,10000)\) \\
         \(\omega_{b}\) & \(\mathcal{N}(6.1,0.2)\) & \(\mathcal{N}(6.1,0.2)\) \\
         \(\omega_{c}\) & \(\mathcal{N}(6.0,0.2)\) & \(\mathcal{N}(6.0,0.2)\) \\
         \(e_{b}\) & \(\mathcal{U}(0,0.01)\) & \(\mathcal{N}(0.3755,0.08)\) \\
         \(e_{c}\) & \(\mathcal{U}(0,0.01)\) & \(\mathcal{N}(0.775,0.07)\) \\
         \(V_{\rm sys}\,[{\rm m\,s^{-1}}]\) & \(\mathcal{U}(0,10000)\) & \(\mathcal{U}(0,10000)\) \\
         \({\rm Jitter}\,[{\rm m\,s^{-1}}]\) & \(\mathcal{MLU}(0.1,100)\) & \(\mathcal{MLU}(0.1,100)\)\\
         \hline
    \end{tabular}
    \label{tab:kima_priors}
\end{table}

This analysis samples combinations of two planets that are consistent with the radial velocity data and with the priors informed from the photometry. The posterior distributions explored correspond to all the possible planets that could exist and we can conclude that planet masses in excess of these distributions are inconsistent with the radial velocity data and can be rejected allowing us to place an upper limit on the mass of both planets in a single analysis. This method is similar to the detection limit method described in \cite{standing2022}, also using {\sc kima}.

\begin{figure*}
    \centering
    \includegraphics[width=1.5\columnwidth]{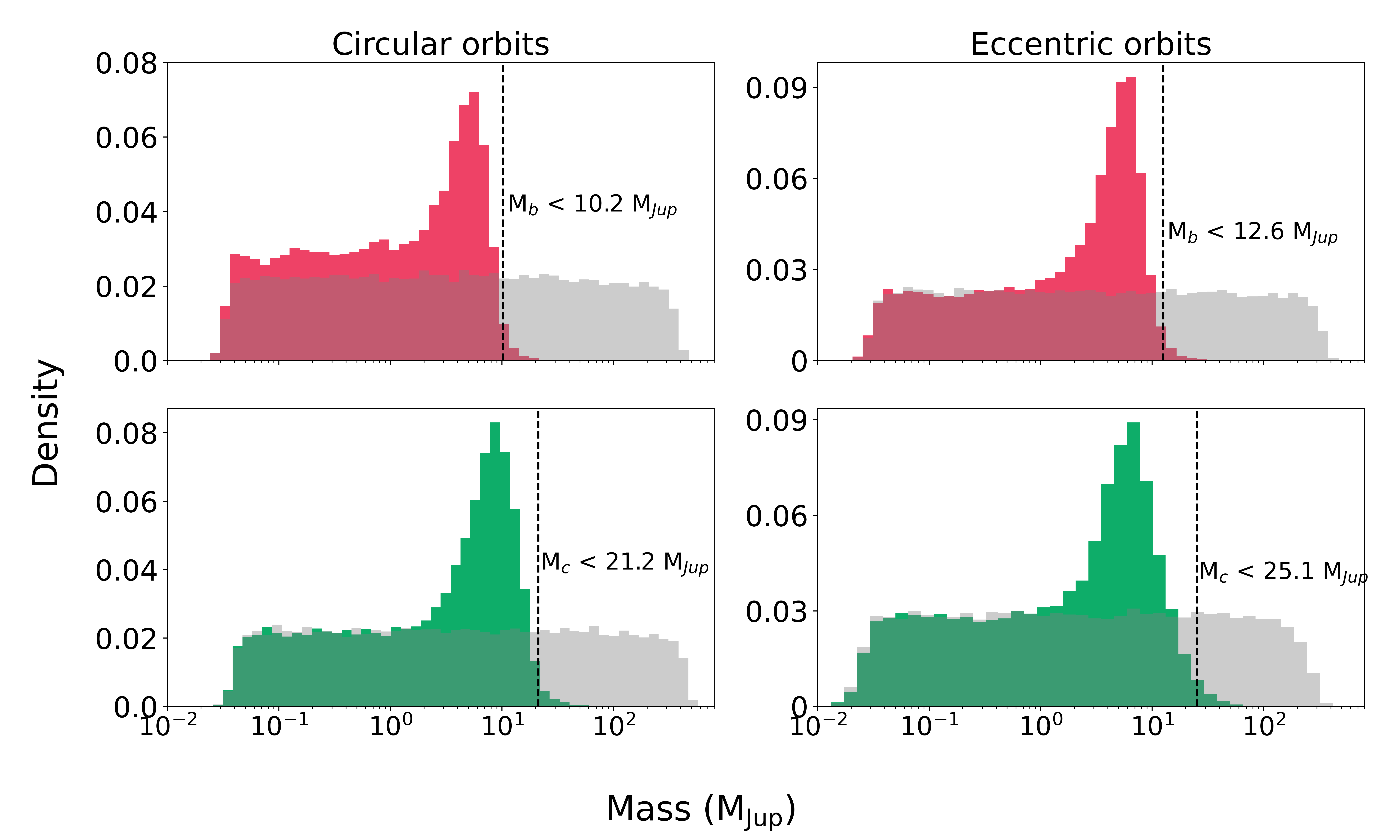}
    \caption{Posterior distribution \review{ from {\sc kima} analysis of radial velocity data }in colour for the mass of each planet assuming circular orbits (left) or with eccentric orbits (right). The prior distributions are represented in grey and in each case the vertical dashed line shows the posterior \(99^{\rm th}\) percentile that we use as our upper limit on the planet mass.}
    \label{fig:hists}
\end{figure*}

The posterior distribution in mass for each planet for each of the two analyses are shown in Figure~\ref{fig:hists} (along with the prior distributions). We find that the 99\% confidence mass upper limits assuming circular orbits are \(M_{\rm b} < 10.2\,M_{\rm Jup}\) and \(M_{\rm c} < 21.2\,M_{\rm Jup}\), and assuming eccentric orbits the upper limits are \(M_{\rm b} < 12.6\,M_{\rm Jup}\) and \(M_{\rm c} < 25.1\,M_{\rm Jup}\).

\begin{figure}
    \centering
    \includegraphics[width=\columnwidth]{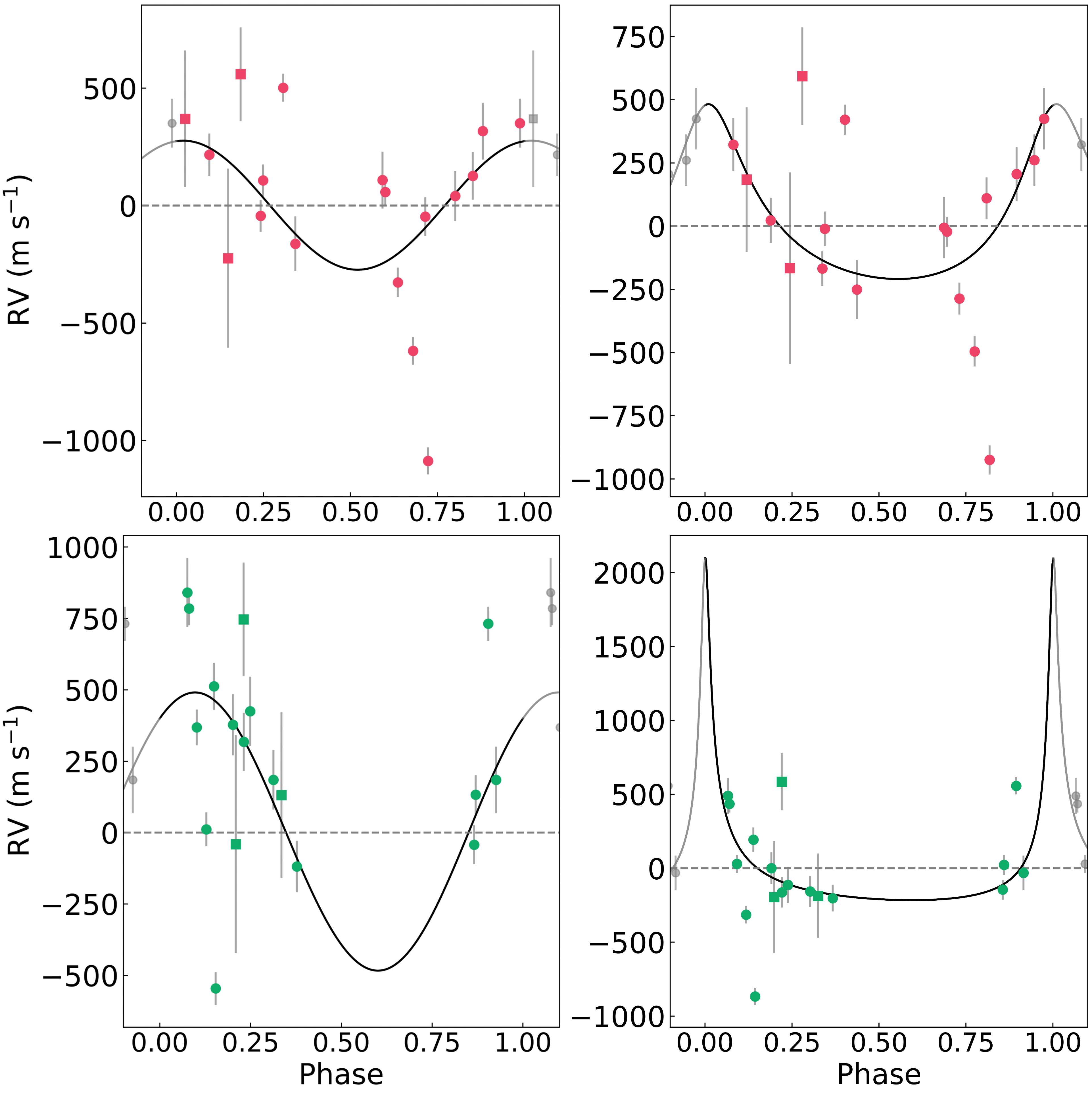}
    \caption{\review{Phased radial velocity data along with models drawn from the {\sc kima} analyses. The two upper panels are for planet b while the bottom panels are planet c. Left-hand panels are circular models and right-hand panels are eccentric. CORALIE measurements are plotted as circles and FEROS as squares.}}
    \label{fig:rv_phased}
\end{figure}

\review{Figure \ref{fig:rv_phased} shows the radial velocity data and a model drawn from each of the circular and eccentric {\sc kima} analyses. These are not a detected radial velocity signature, simply a well-fitting model near the mass limit that we can impose from the data. The circular model has \(M_b = 8.3,\,M_c = 17.5\,M_{\rm Jup}\), the eccentric model has \(M_b = 9.6,\,M_c = 24.0\,M_{\rm Jup}\).}

\subsection{Stability mass constraints}
\label{sec:nepdes}

For giant planets, the spacing between the two planets close to the 5:3 mean motion resonance is very tight. 
For planets on circular orbits, the system becomes Hill unstable (meaning that collisions become possible) if 
\review{their separation in semi-major axis ${a_c-a_b< 2\sqrt{3}R_\mathrm{H}}$ \citep{Gladman1993} where} 
\begin{equation}
R_\mathrm{H} = \frac{a_b+a_c}{2}\left(\frac{m_b+m_c}{3M_\star}\right)^{1/3},
\end{equation}
\review{is the mutual Hill radius. Solving for the known separation between the orbits, we obtain a constraint on}
the total planet mass ${m_b+m_c \leq 2.78 M_{\rm J} (M_\star/M_\odot)}$ . 
The maximum stable mass would be smaller for eccentric configurations \citep{Petit2018}. The orbital stability thus places more stringent constraints onto the planet masses than the radial velocities alone.

\begin{figure}
    \centering
    \includegraphics[width=\linewidth]{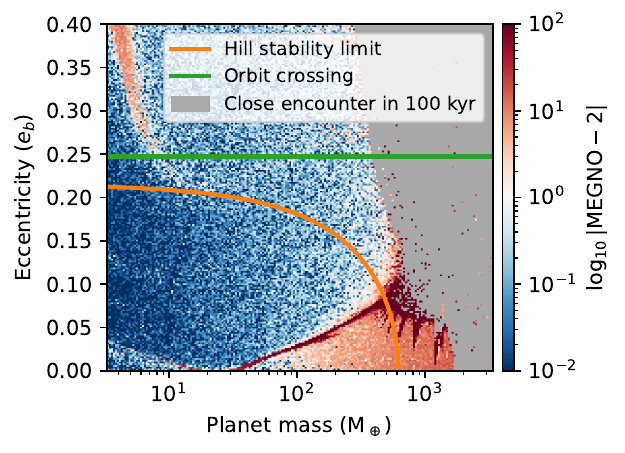}
    \caption{\review{Stability map for systems of two equal mass planet at the period ratio of TOI-791 and planet b eccentricity. Planet c is initialized on a coplanar circular orbit. Systems are integrated for 100 kyr. A MEGNO around 2 indicates a regular system. Grey points indicates simulations that were stopped due to close encounters. We overplot the Hill stability limit \citep{Petit2018} as well as the orbit crossing line.\label{fig:stabmap}}}
\end{figure}

\review{We explore the stability of eccentric configurations at lower masses by performing a stability map analysis. We initialize the system on a grid in mass and eccentricity. As shown by \cite{Hadden2019}, the dynamics are mostly invariant if the total planet mass and relative eccentricity (the difference between the eccentricity vectors) are unchanged.
We thus initialize planet c on a circular orbit while changing planet b eccentricity and choose planets of equal masses.
Due to the proximity to the resonance, we considered initial configurations for which the system is at the resonance center\footnote{We set the mean longitudes of both planets to 0 and choose the longitude of the pericenter to be $\pi/2$ such that the angle $5\lambda_c - 3\lambda_b -2\varpi_b = \pi$ is close to the libration center.}. 
These configurations are the likeliest to be stable as the planets are protected by the resonance locking.
We compute the MEGNO as a chaos indicator over 100 kyr using the \textsc{rebound} integrator library \citep{Rein2012,Rein2015}.
We show the map on Figure \ref{fig:stabmap}. 
We retrieve that masses beyond the Hill circular stability limit are unstable.
For eccentric systems, the resonance protects the systems even beyond the Hill stability limit and orbit crossing. A similar map done with systems initialized outside the resonance closely follows the Hill stability limit constraint.}

\subsection{\review{Transit timing mass constraints}}
\label{sec:ttvanalysis}

\review{We analyse the transit timing variations (TTV) detected for both planets. TOI-791 is expected to show large variations due to the proximity to the resonance.
Indeed TTVs amplitude scale with the planet orbital period which makes TTV modeling an unusually precise method to derive masses at longer periods \citep{Leleu2023}.
A counterpart to the larger amplitude signal is an increase in the period of the TTVs, which in the case of such a resonant configuration exceeds several decades.
Indeed, an integration of our best fit (see below) show a periodic signal of an amplitude larger than a day over a period of 80 years.}

\review{The detected signal is therefore not caused by the resonant contribution to the TTV \citep{Nesvorny2016}, but is instead due to the "chopping" term \citep{Deck2015a} caused by the synodic encounters.
This term has a short period given by the synodic period $P_\mathrm{syn} = (P_b^{-1}-P_c^{-1})$ = 348.6 d.
In typical transiting systems, the chopping effect is usually not detectable given its low-amplitude (of the order of the minute for super-Earths around 10 days).
However, at longer periods, this signal can become the dominant one given the relative shorter baseline (in terms of number of transits).
Unlike the usual TTV signal that presents a degeneracy in mass and eccentricity unless the full TTV period is well resolved \citep{Leleu2023}, the chopping term has a finite amplitude for circular orbit which allows to lift the degeneracy.}

\review{We fit the transit times assuming a coplanar configuration (the low mutual inclination has no impact on the TTVs) using the package \textsc{jnkepler} \citep{Masuda2024}, which computes the transit times gradient with respect to the initial orbital elements.
We then sample around the distribution around the likelihood maximum using the Hamiltonian Monte-Carlo (HMC) sampler implemented in the library \textsc{numpyro} \citep{phan2019,bingham2019}.
We initialized 4 independent chains from the MAP model and evolved them for 2500 steps, discarding 500 steps as burn-in.
The intra-chain autocorrelation indicated no fewer than 1000 independent samples for each parameter; the inter-chain Gelman-Rubin statistic $\hat{R}$ was 1.02 or lower indicating convergence.
We report our best fit result in Table~\ref{tab:ttvmcmc}.
The modeled TTVs are plotted for both planets in Figure \ref{fig:ttvmodel}.
A corner plot showing the parameter correlations is provided in the Appendix \ref{app:TTVcorner}\footnote{\review{The notebooks used to carry out the TTV analysis can be found at \href{https://github.com/acpetit/TOI-791}{https://github.com/acpetit/TOI-791}}}.
}

\begin{figure*}
    \centering
    \includegraphics[width=0.4\linewidth]{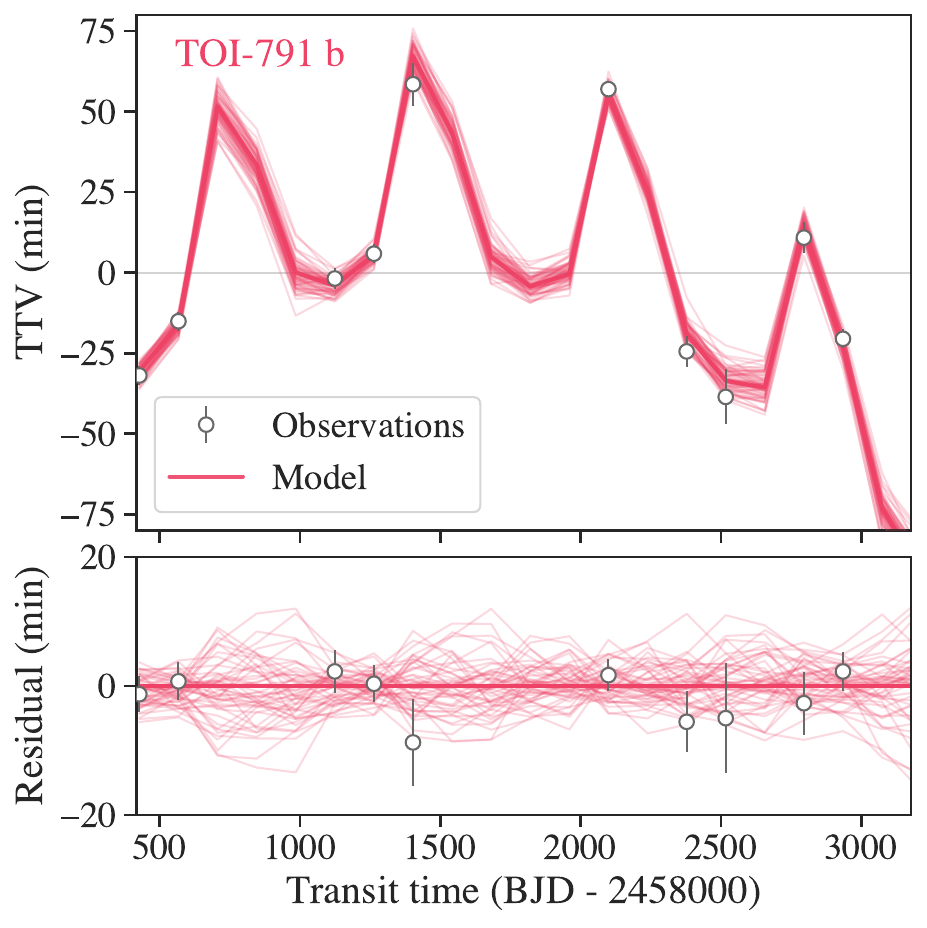}
    \includegraphics[width=0.4\linewidth]{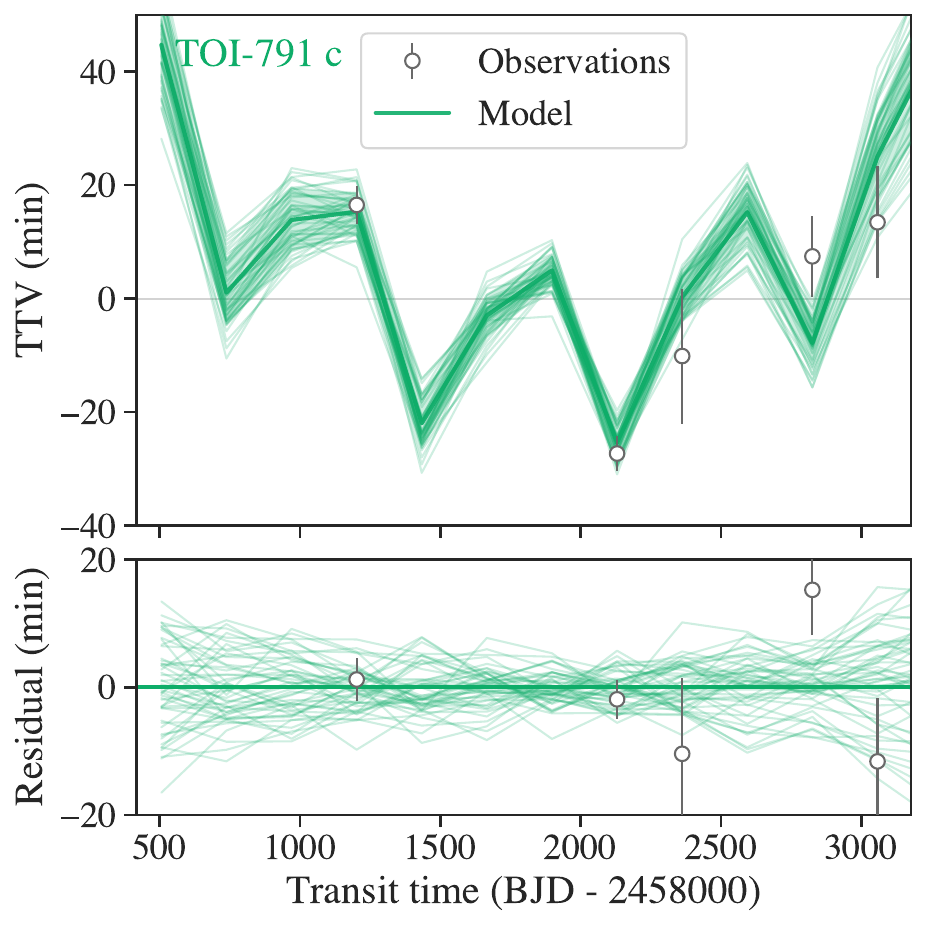}
    \caption{\review{Modelled TTVs for planet b (left) and c (right) as well as their residuals for the best fit (thick line) as well as 30 samples from the posterior distribution. The ribon corresponds to the standard deviation around the baset fit value.\label{fig:ttvmodel}}}
\end{figure*}

\begin{table}
    \centering
    \caption{\review{Prior and result from the posterior sampling from the TTV analysis. For the eccentricities we only provide upper limit estimate since the sampling of the data is not sufficient to characterize enough the orbital configuration. We give the median value as well as errors corresponding to the one $\sigma$ uncertainties for the bayesian highest density  interval. The orbitals elements are given at BJD 2458417.62 and as such the osculating period provided here differs from the mean period obtained from the Keplerian transit fitting. Masses are derived from the planet-to-star mass ratio, assuming a stellar mass drawn from a normal distribution $\mathcal{N}(1.28,0.08)\ M_\odot$.}}
    \begin{tabular}{ll
        r}
        \midrule
         Parameter & Prior & Fitted parameters\\
         \midrule
\(m_{b}/M_s\)  & \(\mathcal{U}(10^{-7},2\times10^{-3})\) &$2.25^{+0.415}_{-0.349}\times 10^{-5}$ \\
\(P_{b}\) [days] & \(\mathcal{U}(139.3,0.1)\) & $139.297^{+0.003}_{-0.002}$ \\
\(T_{b}\) [BJD] & \(\mathcal{U}(430,450)\) &$2458427.623^{+0.002}_{-0.002}$ \\
\(e_b\cos(\omega_{b}) \) & Uniform in $e_b$ &$-0.007^{+0.026}_{-0.035}$ \\
\(e_b\sin(\omega_{b}) \) & Uniform in $e_b$ &$-0.007^{+0.023}_{-0.023}$ \\
\(m_{c}/M_s\)  &\(\mathcal{U}(10^{-7},2\times10^{-3})\)  &$4.39^{+0.433}_{-0.409}\times 10^{-5}$ \\
\(P_{c}\) [days] & \(\mathcal{U}(232.0,0.1)\) &$232.019^{+0.004}_{-0.004}$ \\
\(T_{c}\) [BJD] & \(\mathcal{U}(180,225)\) &$2459201.989^{+0.01}_{-0.013}$ \\
\(e_c\cos(\omega_{c}) \) & Uniform in $e_c$ &$0.018^{+0.022}_{-0.026}$ \\
\(e_c\sin(\omega_{c}) \) & Uniform in $e_c$ &$0.001^{+0.024}_{-0.018}$ \\
        \midrule
        \multicolumn{3}{c}{{ \bf Derived Parameters}}        \\ 
        \midrule    
\(m_{b}\) [\(\mathrm{M}_\oplus\)] &  &$9.5^{+1.64}_{-1.81}$ \\
\(m_{c}\) [\(\mathrm{M}_\oplus\)] &  &$18.6^{+2.24}_{-2.09}$ \\
\(e_b\) & &$0.032^{+0.014}_{-0.032}$ \\
\(e_c\) & &$0.028^{+0.013}_{-0.028}$ \\
\(|Z|\)& &$0.02^{+0.003}_{-0.003}$ \\
\(\Delta\omega\) [rad]  & &$-1.85^{+0.925}_{-1.29}$ \\
\(\lambda_b\) [rad] & & $1.134^{+0.071}_{-0.052}$ \\
\(\lambda_c\) [rad] & & $-0.857^{+0.051}_{-0.045}$ \\
\(\rho_b\) [g.cm$^{-3}$] & & $0.037^{+0.009}_{-0.007}$ \\
\(\rho_c\) [g.cm$^{-3}$] & & $0.047^{+0.008}_{-0.007}$ \\
        \midrule
    \end{tabular}
    \label{tab:ttvmcmc}
\end{table}

\review{The best fit result is obtained for very low masses for both planets compared to their radii. We find that TOI-791 b (respectively c) has a mass of $9.5^{+1.64}_{-1.81}\ \Me$ (resp. \(18.6^{+2.24}_{-2.09}\ \Me\)).
While surprising, the low mass appears robust as we obtain consistent results by resampling assuming different priors in eccentricity and masses as prescribed by \cite{Leleu2023}.
We also ran a similar TTV analysis, assuming the transit timing errors were three times larger, leading to a similar mass range.
}

\review{
The TTV fit is underconstrained since the observing baseline is only able to sample the chopping signal.
Additionally, due to the long orbital periods, we currently do not have consecutive transits covering an entire chopping period, which limits the constraints we can obtain on the eccentricities.
The TTV amplitude depends at first-order in the difference of the eccentricity vector which in complex notation can be written \citep{Hadden2019}}
\begin{equation}
    Z = \frac{e_b e^{\iota \omega_b} - (P_c/P_b)^{0.55} e_c e^{\iota \omega_c}}{\sqrt{1+(P_c/P_b)^{1.1}}}.
    \label{eq:Z}
\end{equation}
\review{
We see that $Z$ is much better constrained than the individual eccentricities and as such we report its value as the true observable.
We observe a slight mass-eccentricity degeneracy in our posterior but it is very limited as the signal also contains terms independent of the eccentricity \citep{Deck2015a}.}

\begin{figure}
    \centering
    \includegraphics[width=\columnwidth]{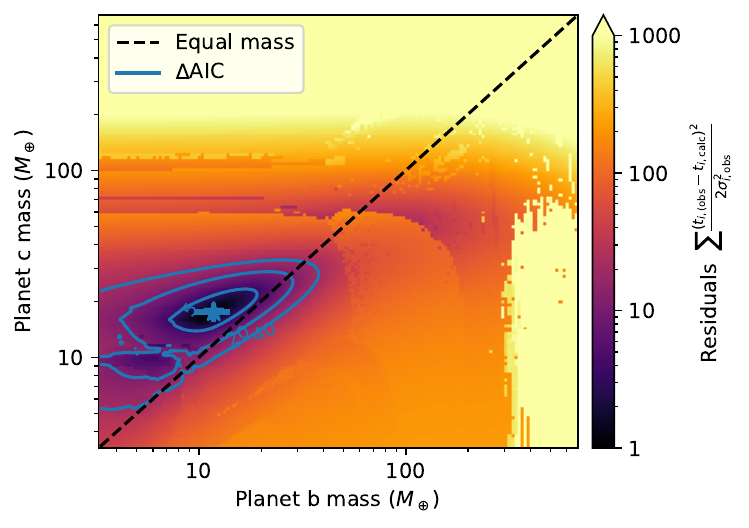}
    \caption{TTVs residuals for the best orbital solution at fixed masses for planet b and c. The level curves labelled 5, 20, 40, correspond to differences in AIC to the best fit solution. Values beyond 40 are strongly ruled out, being half a billion time less likely than the best fit. The star indicate the best fit along with one $\sigma$ error bars.}
    \label{fig:chisq}
\end{figure}

We also tested the robustness of our mass constraints, optimizing the orbital solution at a given mass for both planets to minimize the TTVs residuals.
We initialise a grid of masses logarithmically spaced from \(3 \ \Me\) to \(600\ \Me\) and for each of these point we find the best fit solution by minimising the residuals fixing the masses but varying freely the other parameters \((P,T,e\cos\omega,e\sin\omega)\).
We plot the residual $\chi^2$ map on Fig. \ref{fig:chisq}, where the levels curves show equal values of the difference of Akaike Information Criterion (AIC) to the best fit solution.
Planet c is consistently more massive than planet b where acceptable solutions are found.
Values beyond 40 indicate solutions at least $5\times10^{8}$ less likely than the best fit.
We rule out masses for TOI-791 b above $40\ \Me$ and above $35\ \Me$ for TOI-791 c.
The higher upper bound for planet b is due to the limited number of observed transits for planet c.
As the system will keep being observed, these constraints will be refined.

We also tested the robustness of the fit to systematic errors in the transit times obtained from the ground. For each of the ground observations we draw a systematic error from a normal distribution with a standard deviation of 20 minutes. We then fitted these biased transit times, sampling the posterior using the same procedure as above. We perfom 10 realisations of these biased fit. While not an indication of the accuracy of the transit times we obtained in Section \ref{sec:analysis}, all ten posterior distributions derived from the biased transit times have a worse mean reduced $\chi^2$ (we get $\chi^2 = 10.5\pm 1.6$) than our nominal fit for which $\chi^2=4.5$.
In all ten realisations we get masses estimates within a factor 2 of our reported best fit with many cases showing even lower masses as the best fit.
This test reinforces our confidence in the low-mass nature of planets b and c.
While the exact masses of the planet may still change as more data is acquired on the system, the upper limits at  $40\ \Me$ for TOI-791 b and $35\ \Me$ for TOI-791 c are robust to systematic biases in the transit timing measurements

\section{Discussion}
\label{sec:discussion}



TOI-791 hosts at least two transiting warm giant planets \review{with very low masses, that are likely locked in a 5:3 resonance}. Systems with multiple transiting giants are rare across all orbital periods; TOI-791 is only the $9^{\rm th}$ system known to have more than one transiting giant ($>0.7~R_{\rm J}$) planet. 
All transiting giants in systems that contain more than one giant planet, including TOI-791~b and c, present transit timing variations. This might indicate that dynamical interactions between giants could be a prerequisite for orbital stability when multiple giant planets are present in compact planetary systems. 

\subsection{\review{The extremely low densities of TOI-791 b and c}}

\begin{figure*}
    \centering
    \includegraphics[width=0.75\textwidth]{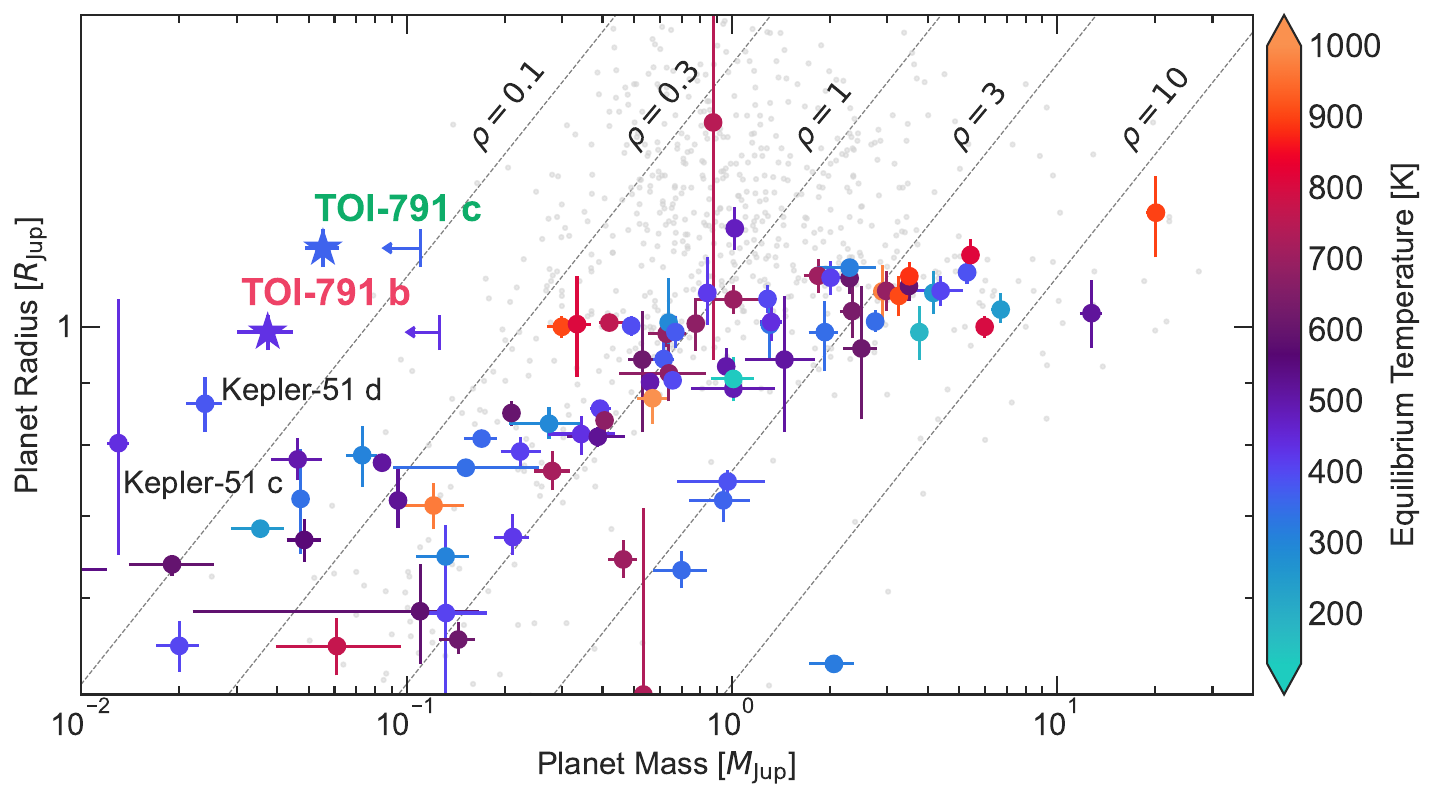}
    \caption{\review{TOI-791 b and c in mass-radius space. The circles are giant planets with orbital periods $>20 \rm~days$ coloured according to equilibrium temperature. In the background, the grey dots are the remainder of the giant planet population (P$\leq$20 days). TOI-791~b and c are indicated by the two labeled stars. We also indicate the mass upper limits from the TTV analysis. The positions of Kepler-51~c and d are shown for comparison. }}
    \label{fig:mass-radius}
\end{figure*}

\review{Our mass constraints from TTVs indicate that TOI-791 b and c are among the lowest density planets detected to date. In Figure \ref{fig:mass-radius} we present a mass-radius plot highlighting the positions of both planets compared to other giant planets with orbital periods longer than 20 days.} TOI-791~c has a larger radius than most of the planets in this parameter space, so a low density is expected. \review{We calculate densities of $\rho_{\rm b}=0.038\pm0.008 \rm ~g~cm^{-3}$ and $\rho_{\rm c}=0.047\pm0.006 \rm ~g~cm^{-3}$ for TOI-791~b and c respectively, making them the largest planets discovered so far with densities lower than $0.05\rm~ g~cm^{-3}$.}

\review{The only planets occupying a comparable parameter space in size, density and orbital period are Kepler-51 b, c and d \citep{Steffen2013,Masuda2014}, a system of super-puff planets. A fourth planet, Kepler-51~e, was recently detected by via transit timing variations \citep{Masuda2024}. While this planet also has a very low mass, as it does not transit it is as yet unclear whether this is a fourth super-puff planet. The similarities between these dynamically interacting siblings and the planets of TOI-791 suggest that the latter may also be super-puffs. However, Kepler-51 is known to be a relatively young system meaning that the planets could still be contracting. Measuring an age for TOI-791 will therefore be crucial to further characterise the system.} Another well studied super-puff is HIP 41378~f \citep{Santerne2019,Bryant2021,Grouffal2025, Lu2025}, a $R_p = 0.82\pm 0.01~\rm R_{J}$ planet with a mass of $12\pm3~\rm M_\oplus$, giving the planet a density of $0.09\pm0.02 \rm ~g~cm^{-3}$. \cite{Santerne2019} highlights that the anomalously low density of this planet is at odds with its radius, which is comparable to Saturn.

\review{The nature of super-puff planets is not well understood. One theory is that they have massive H/He atmospheres accounting for $>20\%$ of their total masses \citep[e.g.][]{Lopez2014}, which requires that they form in the outer regions of protoplanetary discs where large amounts of gas can accrete onto solid cores and cool rapidly \citep{Lee2016}. Another suggestion is that so-called super-puffs are planets of normal density with optically thick rings viewed face-on \citep[e.g.][]{Akinsanmi2020, Ohno2022}. The challenge with this explanation for systems containing multiple super-puffs is that all planets would have to contain similar ring systems, or else have different explanations for their similar low densities \citep{Lammers2024}. For TOI-791, like Kepler-51, the degeneracy can only be lifted with continued observations and further characterisation of the system.}





\subsection{\review{Is the system locked in the 5:3 resonance?}}
\label{sec:53res}

\review{The planets sit extremely close to the 5:3 MMR with a fractional distance to the resonance of $\Delta = 3/5P_c/P_b-1=0.07\%$. They are therefore likely locked in the resonance since even an eccentricity as low as 0.01 would place the system either in the resonance or the chaotic zone around (where it could not remain for long). 
However, confirming the resonant configuration requires a good measurement of the eccentricty and more importantly the longitude of pericenters since the resonant angle is the argument of the complex number $\bar{Z}^2e^{\iota(5\lambda_c-3\lambda_b)}$ \citep{Hadden2019}. Indeed, when both planets are massive and eccentric, the classical resonant angles can circulate while the system is trapped in the resonance as only the difference of eccentricities contributes to the resonant dynamics.}

\begin{figure}
    \centering
    \includegraphics[width=\linewidth]{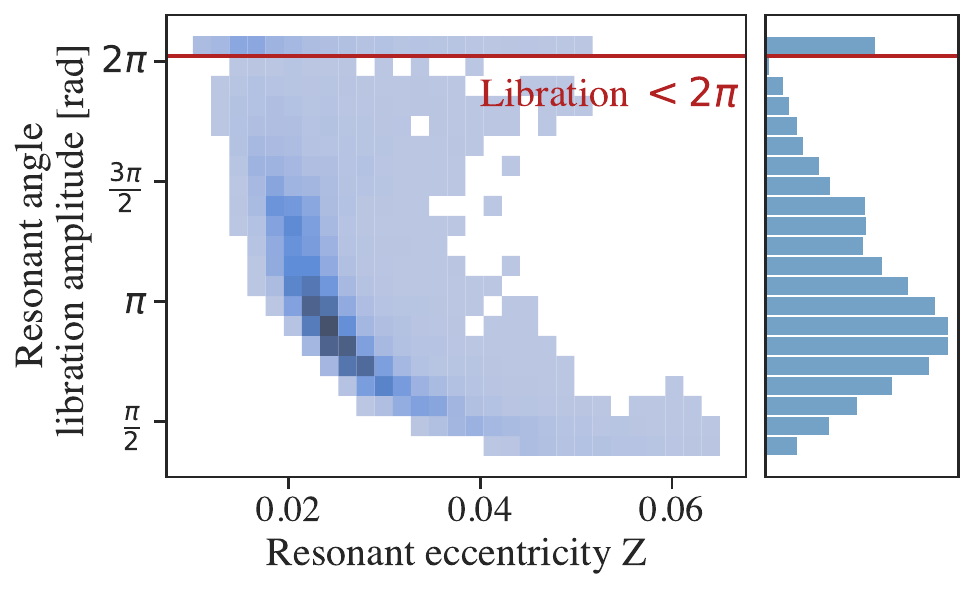}
    \caption{\review{Maximum resonant angle libration amplitude as a function of the resonant eccentricity over 2000 years for initial conditions sampled from the TTV fit posterior distribution. Values above $2\pi$ are circulating and are considered out of resonance. These non-resonant solutions mostly appear at low eccentricity. Over 95\% of the initial conditions are trapped in the resonance.} }
    \label{fig:53res}
\end{figure}

\review{We integrate over 2000 years the initial conditions obtained from our TTV analysis and measure the amplitude of the resonant angle.
We plot the distribution of this amplitude in Figure \ref{fig:53res} as a function of the resonant eccentricity $Z$ (eq. \ref{eq:Z}), which plays a role in the resonant dynamics and is the only component well constrained in a TTV analysis. We find that almost all integrated solutions lie in a resonant configuration (95\% of them have a finite libr amplitude). While the exact configuration would need more observations to be confirmed, there are strong hints that the system is indeed trapped in the 5:3 MMR.   }

\subsection{\review{Prospects for improved precision on planet masses}}
\label{sec:mass}


\review{The mass measurements of TOI-791 b and c cannot be improved by further radial velocity investigation. Even the most conservative upper limits from our TTV analysis ($<40~\rm M_{\oplus}$) would result in RV semi-amplitudes $<10~\rm m s^{-1}$. The detection of such a signal would be impossible given the limitations on this star, namely the rotationally induced line broadening caused by the high $v~\sin i$, and the limited number of spectral lines due to the high effective temperature of the host. }

\review{Instead, increased precision on the planetary mass measurements can be pursued by continued photometric monitoring to better resolve the chopping signal. The ideal scenario would be the observation of a full chopping cycle (5 transits for planet b, 3 for planet c), which would cover an observing baseline of 700 days. Given the star's observability, this would require a combined observing campaign from space and the ground to catch consecutive transits and thus better constrain the orbital configuration of the system. Unfortunately, TOI-791 lies just outside of the \textit{PLATO} LOPS2 \citep{nascimbeni_plato,eschen_plato} and it is also not within the \textit{CHEOPS} field of regard \citep{benz_cheops}. Therefore \textit{TESS} will continue to be the best space-based instrument to observe TOI-791.}

\subsection{How can we further characterise this system?}

TOI-791~b and c are highly amenable to atmospheric follow-up. Space-based observations by \textit{JWST} would be able to capture the broadband spectral features in their atmospheres via transmission spectroscopy. Warm giant planets, especially those with equilibrium temperatures between 300-600~K, are an intriguing sub-population as carbon-, nitrogen-, and oxygen-bearing species should all be observable \citep{Fortney2020}. Therefore, constraining their chemistry allows for multiple elemental ratios to be measured and these have been proposed as potential tracers of planet formation (e.g., C/O \citet{Oberg2011}; N/O \citet{Turrini2021}).

Additionally, TOI-791 is only the 9$^{\rm th}$ system to be found which contains a pair of transiting giant planets. As both planets were formed from the same protoplanetary disc, they provide an excellent opportunity for intra- and inter-system comparative exoplanetology \citep[e.g.,][]{Havel+2011}. Furthermore, as this observational method is sensitive to the atmospheric scale height, these observations could place independent constraints on the masses of these planets \citep{de_wit_seager_mass}, and the exceptionally high precision of these transit lightcurves could allow measurement of the planets' rotational oblateness \citep{Seager2002, Lammers2024}. TOI-791~b and c could additionally be targets for the upcoming \ariel mission, which will carry out a survey of 1000 exoplanetary atmospheres \citep{tinetti_ariel}.

Finally, TOI-791~b and c are excellent candidates for stellar obliquity measurements due to their deep transits and the large $v~\sin{i}$ of the host star. These investigations can give clues about the formation location and potential migration mechanisms of the planets \citep{Triaud2018, Albrecht2022}. Given the high level of rotational broadening of the star's spectral lines, it will
probably be advantageous to analyze
the Rossiter-McLaughlin (R-M) effect as
a distortion of the line profiles, rather than as an anomalous Doppler shift  \citep[e.g.][]{Cameron2010,Albrecht2022,Brady2023}. Using this method carries the added advantage that \review{not as many spectra are needed out of transit, so partial transits are sufficient, which is highly beneficial given the extremely long duration of the transits}. This would add to the growing list of warm giant planets with measured stellar obliquities \citep[e.g.][]{Rice2022,Dong2024,Wang2024}.


\section{Conclusions}
\label{sec:conc}

In this work we have presented the discovery of a pair of long period ($P_b=139.29931_{-0.00012}^{+0.00011}\rm ~days$ and $P_c=232.01570_{-0.00071}^{+0.00067}\rm ~days$) warm Jupiter-sized planets transiting an F7 star, representing the $9^{\rm th}$ known system containing multiple transiting giant planets. The pair have very similar sizes ($R_b=0.993\pm0.033~\rm R_{Jup}$ and $R_c=1.155\pm0.040~\rm R_{Jup}$), and our stability analysis has indicated that they are likely trapped in a 5:3 mean motion resonance. Further long-term monitoring of the system will shed further light on this as more transits of both planets are observed. 

Analysis of the \review{current series of transit timing variations} has allowed us to \review{measure the masses of both planets: $M_{\rm b}=9.5^{+1.64}_{-1.81}\ {\rm M_{\oplus}}$ and $M_{\rm c}=18.6^{+2.24}_{-2.09}\ {\rm M_{\oplus}}$. We find them to be remarkably low density, comparable to the super-puff planets of the Kepler-51 system. {The mass constraints and upper limits derived in this work are robust with the existing dataset; however, continued long-term monitoring of the system is needed to fully characterise the TTV signal and the architecture of the system.}} At this time, the evidence points to a circular configuration for both planets, but ongoing monitoring of the system will in time reveal whether there are slight eccentricities present. 

We have also demonstrated that both planets are suitable for further characterisation. Their mass \review{measurements} can be \review{refined} in time via photodynamical modeling, with precision improving as more transit timings are measured.

TOI-791 is a fascinating laboratory to investigate a pair of dynamically interacting, multi-transiting, \review{low density} co-natal planets, and future in-depth investigations into both planets will add new pieces to the puzzle of giant planet formation.

\section{List of Affiliations}

$^{1}$ Department of Astrophysics, University of Oxford, Denys Wilkinson Building, Keble Road, Oxford OX1 3RH, UK \\
$^{2}$ Magdalen College, University of Oxford, Oxford OX1 4AU, UK \\
$^{3}$ School of Physics \& Astronomy, University of Birmingham, Edgbaston, Birmingham B15 2TT, United Kingdom \\
$^{4}$ Universit\'{e} C\^ote d'Azur, Observatoire de la C\^ote d'Azur, CNRS, Laboratoire Lagrange, Bd de l'Observatoire, CS 34229, 06304 Nice cedex 4, France \\
$^{5}$ Instituto de Astrof\'isica de Canarias (IAC), Calle V\'ia L\'actea s/n, 38200, La Laguna, Tenerife, Spain \\
$^{6}$ Astrobiology Research Unit, Universit\'e de Li\`ege, 19C All\'ee du 6 Ao\^ut, 4000 Li\`ege, Belgium \\
$^{7}$ Department of Physics and Kavli Institute for Astrophysics and Space Research, Massachusetts Institute of Technology, Cambridge, MA 02139, USA \\
$^{8}$ Millennium Institute for Astrophysics, Santiago, Chile \\
$^{9}$ Facultad de Ingeniería y Ciencias, Universidad Adolfo Ibáñez, Av. Diagonal las Torres 2640, Peñalolén, Santiago, Chile \\
$^{10}$ Data Observatory Foundation, Santiago, Chile \\
$^{11}$ Center for Astrophysics \textbar  Harvard \& Smithsonian, 60 Garden Street, Cambridge, MA 02138, USA \\
$^{12}$ SRON, Space Research Organisation Netherlands, Niels Bohrweg 4, NL-2333 CA, Leiden, The Netherlands \\
$^{13}$ El Sauce Observatory, Coquimbo Province, Chile \\
$^{14}$ Observatoire de Gen\`eve, D\'epartement d’Astronomie, Universit\'e de Gen\`eve, Chemin Pegasi 51b, 1290 Versoix, Switzerland \\
$^{15}$ NASA Ames Research Center, Moffett Field, CA 94035, USA \\
$^{16}$ Campo Catino Astronomical Observatory, Regione Lazio, Guarcino (FR), 03010 Italy \\
$^{17}$ Institute of Space Sciences (ICE, CSIC), Carrer de Can Magrans S/N, Campus UAB, Cerdanyola del Valles, E-08193, Spain \\
$^{18}$ Institut d’Estudis Espacials de Catalunya (IEEC), 08860 Castelldefels (Barcelona), Spain \\
$^{19}$ Thüringer Landessternwarte Tautenburg, Sternwarte 5, D-07778, Tautenburg, Germany \\
$^{20}$ Department of Physics \& Astronomy, Vanderbilt University, 6301 Stevenson Center Ln., Nashville, TN 37235, USA \\
$^{21}$ Hazelwood Observatory, Australia \\
$^{22}$ Perth Exoplanet Survey Telescope, Perth, Western Australia, Australia \\
$^{23}$ Department of Physics, Engineering and Astronomy, Stephen F. Austin State University, 1936 North St, Nacogdoches, TX 75962, USA \\
$^{24}$ Univ. Grenoble Alpes, CNRS, IPAG, F-38000 Grenoble, France \\
$^{25}$ European Space Agency (ESA), European Space Research and Technology Centre (ESTEC), Keplerlaan 1, 2201 AZ Noordwijk, The Netherlands \\
$^{26}$ Sapienza Università di Roma, Piazzale Aldo Moro, 5, 00185, Rome (RM), Italy \\
$^{27}$ Università degli Studi di Roma Tor Vergata, Via Cracovia, 50, 00133, Rome (RM), Italy \\
$^{28}$ INFN Sezione Roma1, Piazzale Aldo Moro, 2, 00185, Rome (RM), Italy \\
$^{29}$ INAF OAC, Via della Scienza, 5, 09047, Selargius (CA), Italy \\
$^{30}$ Space Sciences, Technologies and Astrophysics Research (STAR) Institute, Universit\'e de Li\`ege, All\'ee du 6 Ao\^ut 19C, B-4000 Li\`ege, Belgium \\
$^{31}$ Department of Earth, Atmospheric and Planetary Sciences, Massachusetts Institute of Technology, Cambridge, MA 02139, USA \\
$^{32}$ Department of Aeronautics and Astronautics, MIT, 77 Massachusetts Avenue, Cambridge, MA 02139, USA \\
$^{33}$ European Southern Observatory, Alonso de Córdova 3107, Vitacura, Región Metropolitana, Chile \\
$^{34}$ Landessternwarte, Zentrum für Astronomie der Universtät Heidelberg, Königstuhl 12, 69117 Heidelberg, Germany \\
$^{35}$ Department of Astronomy, Sofia University St Kliment Ohridski, 5 James Bourchier Blvd, BG-1164 Sofia, Bulgaria \\
$^{36}$ SETI Institute, Mountain View, CA 94043 USA/NASA Ames Research Center, Moffett Field, CA 94035 USA \\
$^{37}$ Department of Astrophysical Sciences, Princeton University, Princeton, NJ 08544, USA \\


\section*{Acknowledgements}
\review{The authors would like to thank the reviewers for their comments which allowed us to significantly expand and improve the results from this paper.} {The authors would also like to thank Eric Agol for the helpful discussions on the analysis of the TTVs of this system.}
This work makes use of observations from the ASTEP telescope. ASTEP benefited from the support of the French and Italian polar agencies IPEV and PNRA in the framework of the Concordia station program and from OCA, INSU, Idex UCAJEDI (ANR- 15-IDEX-01) and ESA through the Science Faculty of the European Space Research and Technology Centre (ESTEC). This research also received funding from the European Research Council (ERC) under the European Union's Horizon 2020 research and innovation program (grant agreement No. 803193/BEBOP), from the Science and Technology Facilities Council (STFC; grant No. ST/S00193X/1, ST/W002582/1 and ST/Y001710/1), and from the ERC/UKRI Frontier Research Guarantee programme (CandY/ EP/Z000327/1).
Funding for the TESS mission is provided by NASA's Science Mission Directorate. We acknowledge the use of public TESS data from pipelines at the TESS Science Office and at the TESS Science Processing Operations Center. Resources supporting this work were provided by the NASA High-End Computing (HEC) Program through the NASA Advanced Supercomputing (NAS) Division at Ames Research Center for the production of the SPOC data products.
This research has made use of the Exoplanet Follow-up Observation Program (ExoFOP; DOI: 10.26134/ExoFOP5) website, which is operated by the California Institute of Technology, under contract with the National Aeronautics and Space Administration under the Exoplanet Exploration Program. This paper includes data collected by the TESS mission that are publicly available from the Mikulski Archive for Space Telescopes (MAST). KAC and CNW acknowledge support from the TESS mission via subaward s3449 from MIT.
Some of the observations in this paper made use of the High-Resolution Imaging instrument Zorro and were obtained under Gemini LLP Proposal Number: GN/S-2021A-LP-105. Zorro was funded by the NASA Exoplanet Exploration Program and built at the NASA Ames Research Center by Steve B. Howell, Nic Scott, Elliott P. Horch, and Emmett Quigley. Zorro was mounted on the Gemini South telescope of the international Gemini Observatory, a program of NSF’s OIR Lab, which is managed by the Association of Universities for Research in Astronomy (AURA) under a cooperative agreement with the National Science Foundation. on behalf of the Gemini partnership: the National Science Foundation (United States), National Research Council (Canada), Agencia Nacional de Investigación y Desarrollo (Chile), Ministerio de Ciencia, Tecnología e Innovación (Argentina), Ministério da Ciência, Tecnologia, Inovações e Comunicações (Brazil), and Korea Astronomy and Space Science Institute (Republic of Korea).
The research leading to these results has received funding from  the ARC grant for Concerted Research Actions, financed by the Wallonia-Brussels Federation. TRAPPIST is funded by the Belgian Fund for Scientific Research (Fond National de la Recherche Scientifique, FNRS) under the grant PDR T.0120.21.
MG and EJ are F.R.S.-FNRS Research Directors.
This research has made use of the NASA Exoplanet Archive, which is operated by the California Institute of Technology, under contract with the National Aeronautics and Space Administration under the Exoplanet Exploration Program.
We thank the Swiss National Science Foundation (SNSF) and the Geneva University for their continuous support. 
This work has been carried out within the framework of the National Centre of Competence in Research PlanetS sup-
ported by the Swiss National Science Foundation under grants 51NF40\_182901and 51NF40\_205606.
This work makes use of observations from the LCOGT network. Part of the LCOGT telescope time was granted by NOIRLab through the Mid-Scale Innovations Program (MSIP). MSIP is funded by NSF.
GD acknowledges funding from Magdalen College, Oxford.
A.C.P. has been supported by the French government, through the $ \mathrm{UCA^{J.E.D.I.}} $ Investments in the Future project managed by the National Research Agency (ANR) with the reference number ANR-15-IDEX-01.
CAC acknowledges that this research was carried out at the Jet Propulsion Laboratory, California Institute of Technology, under a contract with NASA (80NM0018D0004).
AVF acknowledges the support of the IOP through the Bell Burnell Graduate Scholarship Fund.
A.J. acknowledges support from ANID -- Millennium  Science  Initiative -- ICN12\_009, AIM23-0001 and from FONDECYT project 1210718.
The postdoctoral fellowship of KB is funded by F.R.S.-FNRS grant T.0109.20 and by the Francqui Foundation.
AS acknowledges support from the European Research Council Consolidator Grant funding scheme (project ASTEROCHRONOMETRY, G.A. n. 772293, http://www.asterochronometry.eu).
AP acknowledges support from the Unidad de Excelencia María de Maeztu CEX2020-001058-M programme and from the Generalitat de Catalunya/CERCA.
MBN acknowledges support from the UK Space Agency.
MS acknowledges financial support from the Swiss National Science Foundation (SNSF) for project 200021\_200726. This project has also been carried out in the frame of the National Centre for Competence in Research PlanetS supported by the SNSF.
A.J. and R.B. acknowledge support from ANID -- Millennium  Science  Initiative -- ICN12\_009 and AIM23-0001.
R.B. acknowledges support from FONDECYT Project 1241963.
A.J.\ acknowledges support from FONDECYT project 1210718.
Funding for KB was provided by the European Union (ERC AdG SUBSTELLAR, GA 101054354).
A.S. postdoctoral fellowship is funded by F.R.S.-FNRS research project ID 40028002 (Detection and Study of Rocky Worlds).

This research made use of Lightkurve, a Python package for Kepler and TESS data analysis (Lightkurve Collaboration, 2018). 
This work made use of Astropy:\footnote{http://www.astropy.org} a community-developed core Python package and an ecosystem of tools and resources for astronomy \citep{astropy:2013, astropy:2018, astropy:2022}. 
This research has made use of the NASA Exoplanet Archive, which is operated by the California Institute of Technology, under contract with the National Aeronautics and Space Administration under the Exoplanet Exploration Program. 
This research made use of \textsf{exoplanet} \citep{exoplanet} and its
dependencies \citep{exoplanet:agol20, exoplanet:arviz, exoplanet:astropy13,
exoplanet:astropy18, exoplanet:luger18, exoplanet:pymc3, exoplanet:theano}.
This research has made use of \textsc{AstroImageJ} \citep{Collins:2017} and {\sc TAPIR} \citep{Jensen:2013}.

\section*{Data Availability}

\tess data products are available via the MAST portal at \url{https://mast.stsci.edu/portal/Mashup/Clients/Mast/Portal.html}. Follow-up photometry and high resolution imaging data for TOI-791 are available on ExoFOP at \url{https://exofop.ipac.caltech.edu/tess/target.php?id=306472057}. These data are freely accessible to ExoFOP members immediately and are publicly available following a one-year proprietary period.



\bibliographystyle{mnras}
\bibliography{TOI-791} 




\appendix

\section{CORALIE \review{and FEROS} radial velocities}

In Table \ref{tab:coralie} we present the 15 RV measurements we collected using CORALIE\review{, and in Table \ref{tab:feros} we present the 3 RV measurements collected with FEROS}. 

\begin{table}
    \centering
    \begin{tabular}{ccc}
    \textbf{BJD} & \textbf{RV \review{($\rm km\,s^{-1}$)}} & \textbf{Error \review{($\rm km\,s^{-1}$)}}\\
    \midrule
         2458804.77709573 & 5.53246 & 0.06796 \\
2458805.74093483 & 5.69562 & 0.06785 \\
2458813.80235570 & 6.19564 & 0.05939 \\
2458854.66732408 & 6.07403 & 0.05915 \\
2458859.64605163 & 5.69218 & 0.06272 \\
2458865.71770650 & 5.39155 & 0.05944 \\
2458871.69900158 & 4.90070 & 0.05742 \\
2458889.62034295 & 5.98041 & 0.10175 \\
2459514.84538964 & 5.59217 & 0.11673 \\
2459549.70801438 & 6.12348 & 0.12120 \\
2459566.74670700 & 5.94574 & 0.08217 \\
2459578.76636795 & 5.95754 & 0.10656 \\
2459589.79074966 & 6.12848 & 0.12115 \\
2459604.66921639 & 5.98416 & 0.10414 \\
2459619.60042030 & 5.65639 & 0.08989 \\
    \end{tabular}
    \caption{RV data collected with CORALIE. }
    \label{tab:coralie}
\end{table}

\begin{table}
    \centering
    \begin{tabular}{ccc}
    \textbf{BJD} & \textbf{RV \review{($\rm km\,s^{-1}$)}} & \textbf{Error \review{($\rm km\,s^{-1}$)}}\\
    \midrule
         2458652.50577&5.8540&0.3695 \\
         2458657.49966&6.5914&0.1749 \\
         2458913.77533&6.1129&0.2744\\
    \end{tabular}\\
    \caption{RV data collected with FEROS. }
    \label{tab:feros}
\end{table}

\section{Measured transit times}

\begin{table}
    \centering
    \begin{tabular}{ccc}
    \textbf{Planet} & \textbf{Midtime [BJD]} & \textbf{Error}\\
    \midrule\midrule
         b & 2458427.62210141 & 0.00195261  \\
         b & 2458566.93358664 & 0.0020498  \\
         b & 2459124.14202378 & 0.00232475  \\
         b & 2459263.44717619 & 0.00199896 \\
         b & 2459402.78350626 & 0.00464576 \\
         b & 2460099.28145936 & 0.00172244 \\
         b & 2460377.82453042 & 0.00330275 \\
         b & 2460517.11452191 & 0.00592322 \\
         b & 2460795.74846824 & 0.0034036 \\
         b & 2460935.02646272 & 0.00207694 \\
         \midrule
         c & 2459201.96844372 & 0.00231274 \\
         c & 2460130.01643396 & 0.00211466 \\
         c & 2460362.04797325 & 0.00826268 \\
         c & 2460826.0993888 & 0.00492143 \\
         c & 2461058.12316399 & 0.00686396 \\
    \end{tabular}\\
    \caption{Measured transit midtimes for both planets}
    \label{tab:ttvs}
\end{table}

\section{TTV analysis corner plot}
\label{app:TTVcorner}
\begin{figure*}
    \centering
    \includegraphics[width=\linewidth]{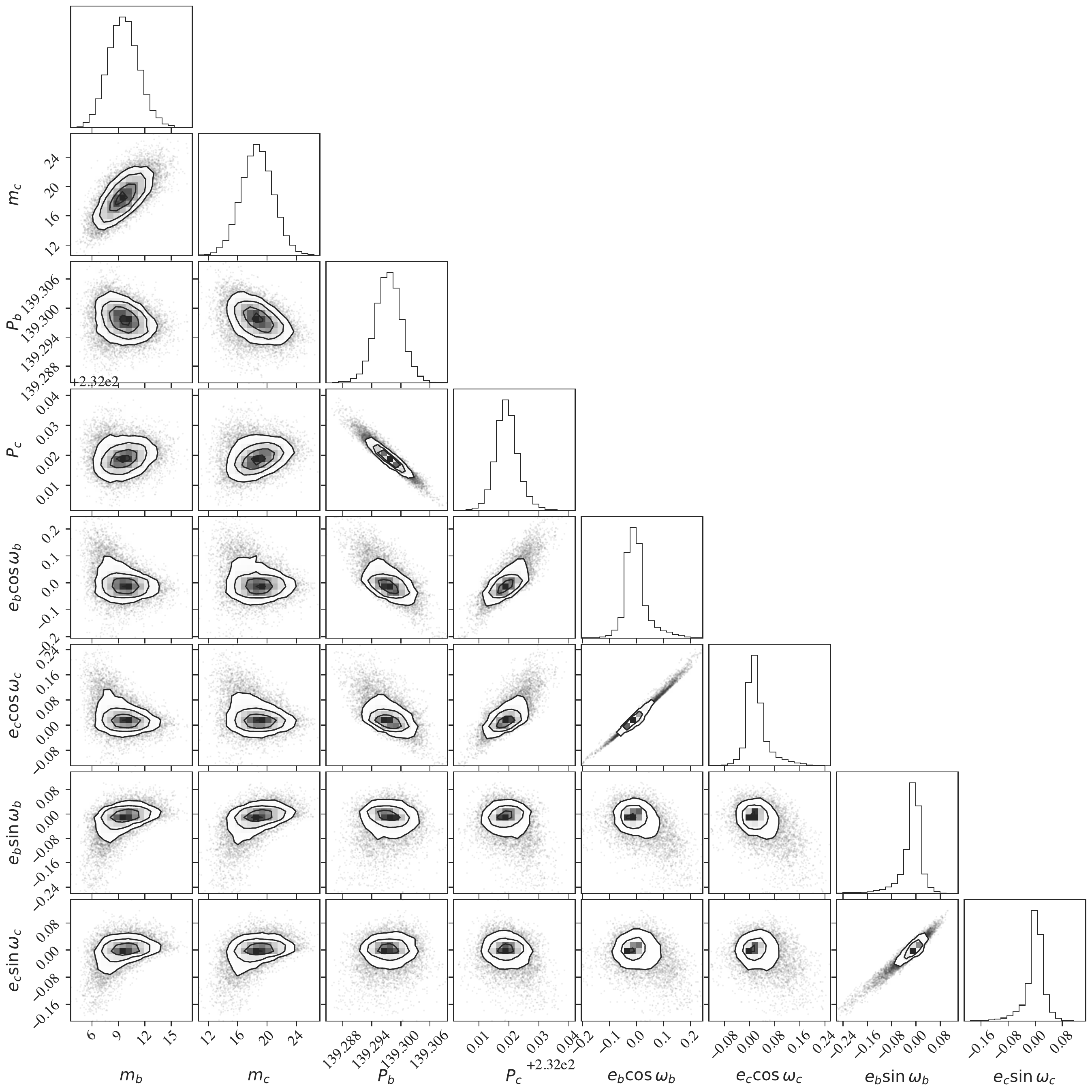}
    \caption{Corner plot for the MCMC samples of the TTV analysis described in section \ref{sec:ttvanalysis}. We highlight the strong correlation between the eccentricity components. Masses are given in $M_\oplus$ and period in days. The transit times were removed from the plots to reduce clutterness as they show not much meaningful information. We also report the masses rather than the planet-to-star mass ratio which were the variables we sampled the distribution with.}
\end{figure*}



\bsp	
\label{lastpage}
\end{document}